\documentstyle[preprint]{jpsj}
\newread\epsffilein    
\newif\ifepsffileok    
\newif\ifepsfbbfound   
\newif\ifepsfverbose   
\newif\ifepsfdraft     
\newdimen\epsfxsize    
\newdimen\epsfysize    
\newdimen\epsftsize    
\newdimen\epsfrsize    
\newdimen\epsftmp      
\newdimen\pspoints     
\def\epsfbox#1{\global\def\epsfllx{72}\global\def\epsflly{72}%
   \global\def\epsfurx{540}\global\def\epsfury{720}%
   \def\lbracket{[}\def\testit{#1}\ifx\testit\lbracket
   \let\next=\epsfgetlitbb\else\let\next=\epsfnormal\fi\next{#1}}%
\def\epsfgetlitbb#1#2 #3 #4 #5]#6{\epsfgrab #2 #3 #4 #5 .\\%
   \epsfsetgraph{#6}}%
\def\epsfnormal#1{\epsfgetbb{#1}\epsfsetgraph{#1}}%
\def\epsfgetbb#1{%
\openin\epsffilein=#1
\ifeof\epsffilein\errmessage{I couldn't open #1, will ignore it}\else
   {\epsffileoktrue \chardef\other=12
    \def\do##1{\catcode`##1=\other}\dospecials \catcode`\ =10
    \loop
       \read\epsffilein to \epsffileline
       \ifeof\epsffilein\epsffileokfalse\else
          \expandafter\epsfaux\epsffileline:. \\%
       \fi
   \ifepsffileok\repeat
   \ifepsfbbfound\else
    \ifepsfverbose\message{No bounding box comment in #1; using defaults}\fi\fi
   }\closein\epsffilein\fi}%
\def\epsfclipoff{\def\epsfclipstring{\ifepsfdraft\space clip\fi}}%
\epsfclipoff
\def\epsfsetgraph#1{%
   \epsfrsize=\epsfury\pspoints
   \advance\epsfrsize by-\epsflly\pspoints
   \epsftsize=\epsfurx\pspoints
   \advance\epsftsize by-\epsfllx\pspoints
   \epsfxsize\epsfsize\epsftsize\epsfrsize
   \ifnum\epsfxsize=0 \ifnum\epsfysize=0
      \epsfxsize=\epsftsize \epsfysize=\epsfrsize
      \epsfrsize=0pt
     \else\epsftmp=\epsftsize \divide\epsftmp\epsfrsize
       \epsfxsize=\epsfysize \multiply\epsfxsize\epsftmp
       \multiply\epsftmp\epsfrsize \advance\epsftsize-\epsftmp
       \epsftmp=\epsfysize
       \loop \advance\epsftsize\epsftsize \divide\epsftmp 2
       \ifnum\epsftmp>0
          \ifnum\epsftsize<\epsfrsize\else
             \advance\epsftsize-\epsfrsize \advance\epsfxsize\epsftmp \fi
       \repeat
       \epsfrsize=0pt
     \fi
   \else \ifnum\epsfysize=0
     \epsftmp=\epsfrsize \divide\epsftmp\epsftsize
     \epsfysize=\epsfxsize \multiply\epsfysize\epsftmp   
     \multiply\epsftmp\epsftsize \advance\epsfrsize-\epsftmp
     \epsftmp=\epsfxsize
     \loop \advance\epsfrsize\epsfrsize \divide\epsftmp 2
     \ifnum\epsftmp>0
        \ifnum\epsfrsize<\epsftsize\else
           \advance\epsfrsize-\epsftsize \advance\epsfysize\epsftmp \fi
     \repeat
     \epsfrsize=0pt
    \else
     \epsfrsize=\epsfysize
    \fi
   \fi
   \ifepsfverbose\message{#1: width=\the\epsfxsize, height=\the\epsfysize}\fi
   \epsftmp=10\epsfxsize \divide\epsftmp\pspoints
   \vbox to\epsfysize{\vfil\hbox to\epsfxsize{%
      \ifnum\epsfrsize=0\relax
        \includegraphics{\ifepsfdraft}%
      \else
        \epsfrsize=10\epsfysize \divide\epsfrsize\pspoints
        \includegraphics{\ifepsfdraft}%
      \fi
      \hfil}}%
\global\epsfxsize=0pt\global\epsfysize=0pt}%
{\catcode`\%=12 \global\let\epsfpercent=
\long\def\epsfaux#1#2:#3\\{\ifx#1\epsfpercent
   \def\testit{#2}\ifx\testit\epsfbblit
      \epsfgrab #3 . . . \\%
      \epsffileokfalse
      \global\epsfbbfoundtrue
   \fi\else\ifx#1\par\else\epsffileokfalse\fi\fi}%
\def\epsfempty{}%
\def\epsfgrab #1 #2 #3 #4 #5\\{%
\global\def\epsfllx{#1}\ifx\epsfllx\epsfempty
      \epsfgrab #2 #3 #4 #5 .\\\else
   \global\def\epsflly{#2}%
   \global\def\epsfurx{#3}\global\def\epsfury{#4}\fi}%
\def\epsfsize#1#2{\epsfxsize}
\def\runtitle{Charge Ordering in Organic ET Compounds}
\def\runauthor{Hitoshi {\sc Seo}}

\newcommand{\lsim}{\hspace{1mm}
\lower -0.3ex \hbox{$<$} \kern -0.75em \lower 0.7ex \hbox{$\sim$}
\hspace{1mm}}
\newcommand{\gsim}{\hspace{1mm}
\lower -0.3ex \hbox{$>$} \kern -0.75em \lower 0.7ex \hbox{$\sim$}
\hspace{1mm}}

\title{Charge Ordering in Organic ET Compounds}
\author{Hitoshi {\sc Seo}\footnote{E-mail: hseo@swan.issp.u-tokyo.ac.jp}}
\inst{Institute for Solid State Physics, University of Tokyo, 
Minato-ku, Tokyo 106-8666, Japan}

\recdate{}

\abst{The charge ordering phenomena in quasi two-dimensional 
1/4-filled organic compounds (ET)$_2X$ (ET=BEDT-TTF) 
are investigated theoretically
for the $\theta$ and $\alpha$-type structures, 
based on the Hartree approximation for the extended Hubbard models 
with both on-site and intersite Coulomb interactions.
It is found that charge ordered states of stripe-type 
are stabilized for the relevant values of Coulomb energies, 
while the spatial pattern of the stripes sensitively depends on the 
anisotropy of the models. 
By comparing the results of calculations with the experimental facts, 
where the effects of quantum fluctuation is incorporated 
by mapping the stripe-type charge ordered states 
to the $S=1/2$ Heisenberg Hamiltonians, 
the actual charge patterns in the insulating phases of 
$\theta$-(ET)$_2MM'$(SCN)$_4$ and $\alpha$-(ET)$_2$I$_3$ 
are deduced. 
Furthermore, to obtain a unified view 
among the $\theta$, $\alpha$ and $\kappa$-(ET)$_2X$ families, 
the stability of the charge ordered state in competition with 
the dimeric antiferromagnetic state viewed as the Mott insulating state, 
which is typically realized in $\kappa$-type compounds, 
and with the paramagnetic metallic state, is also pursued  
by extracting essential parameters. }
\kword{charge ordering, organic conductors, BEDT-TTF, extended Hubbard model, 
Hartree approximation.}

\begin{document}
\sloppy
\maketitle

\section{Introduction}
\label{intro}
The family of quasi two-dimensional (2D) organic conductors 
(ET)$_2X$ (ET=BEDT-TTF) is known to exhibit 
a variety of interesting electronic properties.\cite{IshiYama} 
Their structure consists of 
alternating layers of anionic $X^{-}$ with the closed shell, 
and cationic ET$^{1/2+}$ 
whose $\pi$-band is 3/4-filled 
(1/4-filled in terms of holes) 
if all the ET molecules are equivalent. 
The variety in physical properties reflects 
that of spatial arrangements of ET molecules in the layer, 
which are classified by greek characters 
such as $\kappa$, $\alpha$, $\theta$, $\beta$, etc., 
together with the rather strong mutual Coulomb interaction 
among $\pi$-electrons.\cite{Review} 
Theoretical studies of Kino and Fukuyama
\cite{Kinoalpha,Kinokappa,Kinointer,Kinopaper} (KF) developed 
a systematic way to understand the diversity in their ground state properties 
by taking into account the explicit anisotropy of the 
transfer integrals between ET molecules for each compound, 
and by mean-field (MF) calculations 
treating the on-site Coulomb interaction $U$ within 
the Hartree-Fock approximation. 

By studying the structure of $\kappa$-(ET)$_2X$ 
where the intradimer transfer integral 
is twice larger than the others,\cite{Oshima}  
and by varying the degree of the dimerization, 
KF have investigated the stability of 
the insulating state with antiferromagnetic (AF) ordering between dimers, 
which is called the dimeric AF state.\cite{Kinokappa}
This state is stable when the dimerization in the ET layer is strong, 
since the electron correlation results in a Mott insulating state 
due to the 1/2-filled $\pi$-band,\cite{Kanoda,McKenzie} 
as has been first argued by Tokura.\cite{TokuraET} 

\begin{figure}
\begin{center}
\leavevmode\epsfysize=10cm
\epsfbox{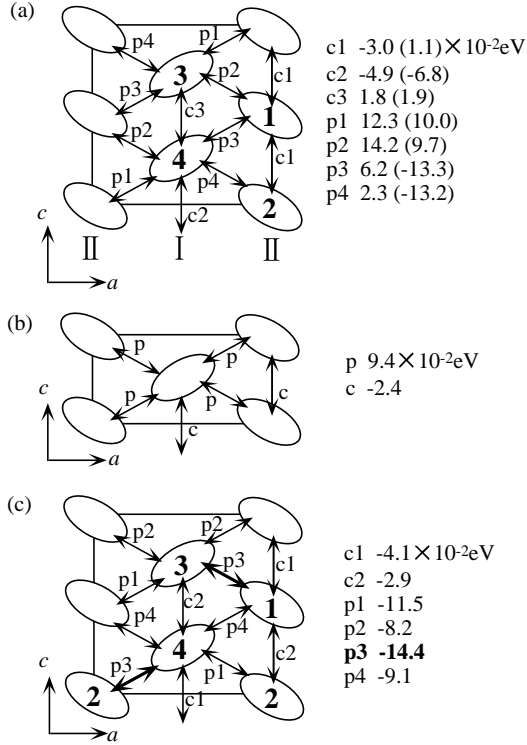}
\end{center}
\caption{Schematic representation of the structures in the donor plane 
for (ET)$_2X$ compounds;  
$\alpha$-type (a), 
$\theta$-type (b) and $\theta_d$-type (c)
(see the text for the notation of $\theta_d$). 
ET molecules are drawn as ellipses, and the intermolecular transfer integrals 
are given by $t = Es$ with $E=-10$ eV, where the overlap integrals 
$s$ are for 
$\alpha$-(ET)$_2$I$_3$\cite{TMori}(a) 
where the indices are modified from those in ref. \ref{TMori} 
as $a1-a3 \rightarrow c1-c3$ and $b1-b4 \rightarrow p1-p4$, 
(those corresponding to $\alpha$-(ET)$_2$NH$_4$Hg(SCN)$_4$ are also shown 
in the parentheses of (a)\cite{MHgSDW}), 
$\theta$-(ET)$_2$CsZn(SCN)$_4$ \cite{Mori}(b) 
and $\theta$-(ET)$_2$RbCo(SCN)$_4$ \cite{Mori}(c).}
\label{structure}
\end{figure}

Another interesting conclusion of KF is that 
the $\alpha$-type compounds 
show tendency toward an insulating state with charge ordering (CO) 
(charge transfer in their notation). \cite{Kinoalpha}
Experimentally, 
$\alpha$-(ET)$_2$I$_3$ shows a metal-insulator (MI) 
transition at $T_{\rm MI}$=135 K\cite{Bender}  
at ambient pressure, 
and by applying pressure this $T_{\rm MI}$ is supressed.\cite{Kajita}
The magnetic susceptibility shows a sharp drop below $T_{\rm MI}$ 
indicating a non-magnetic ground state with a spin gap.\cite{Rothamael} 
A bond length analysis\cite{Heidmann} as well as 
an IR absorbtion experiment\cite{Moldenhauer} in this compound  
suggest charge disproportionation among ET molecules, 
though the suggested spatial patterns of charge disproportionation 
are different from each other. 
The Hartree-Fock calculations by KF for the structure of 
$\alpha$-(ET)$_2$I$_3$ 
show CO between the stacks I and II, 
which are the two crystallographically inequivalent columns 
as shown in Fig. \ref{structure}(a). 
The actual pattern of CO 
as well as the nature of the phase transition in this compound 
is still an open problem. 
Recently some suggestions have been made 
that the ground state of another class of $\alpha$-phase, 
$\alpha$-(ET)$_2M$Hg(SCN)$_4$ for M=K, Rb, and Tl, 
which are isostructural to $\alpha$-(ET)$_2$I$_3$ 
(see Fig. \ref{structure}(a)), 
may also be a charge disproportionated state\cite{MHgCDW}
rather than the so far believed spin-density-wave (SDW) state 
due to the nesting of Fermi surface.\cite{Kinointer,MHgSDW} 

\begin{figure}
\begin{center}
\leavevmode\epsfysize=6cm
\epsfbox{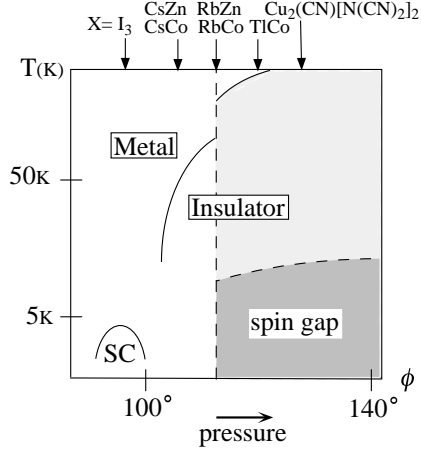}
\end{center}
\caption{Schematic phase diagram of $\theta$-(ET)$_2X$ 
deduced from experiments, 
on the plane of dihedral angle ($\phi$) 
and temperature (T).\cite{Mori,Takahashi} 
The effect of pressure is indicated by the arrow, 
and SC denote the superconducting phase. 
The colored area show the $\theta_d$-phase, 
while the other is the $\theta$-phase. 
The locations of ambient pressure are shown 
for examples of anions $X$, 
where $X=MM'$(SCN)$_4$ ($M$=Rb, Tl, Cs, $M'$=Co, Zn) is abbreviated as $MM'$. 
}
\label{theta_phd}
\end{figure}

Recently such CO has been directly observed in NMR measurements 
by Miyagawa {\it et al}\cite{Miyagawa} and by Chiba {\it et al}\cite{Chiba} 
in a member of another polytype $\theta$-(ET)$_2X$. 
A series of these compounds, 
$\theta$-(ET)$_2MM'$(SCN)$_4$ ($M$=Rb, Tl, Cs, $M'$=Co, Zn), 
has been synthesized by H. Mori {\it et al},\cite{Mori} 
who have proposed that the properties of $\theta$-phase 
can be summarized in one phase diagram, 
where the horizontal axis is the dihedral angle, $\phi$,\cite{note} 
between the molecular planes, 
which is different for each compound. 
It has been argued by Takahashi and Nakumara\cite{Takahashi} 
that a phase diagram, a slightly modified version of ref. \ref{Mori}, 
can be drawn as shown in Fig. \ref{theta_phd}, 
since it has been realized by experiments that 
among the above series, 
located in the region of $100^{\circ}\lsim \phi \lsim 140^{\circ}$, 
there exist compounds with 
two different types of structures, the usually called $\theta$-type one, 
as shown in Fig. \ref{structure}(b), 
and another one with slight dimerization along the $c$-axis,  
as shown in Fig. \ref{structure}(c), 
which we call here $\theta_d$-type to make distinction. 
The electronic properties of these two phases 
are distinct to each other. 
The compounds in the $\theta$-phase 
which exist in the left side of the phase diagram 
($\phi \lsim 110^{\circ}$) 
are metallic at room temperature, 
and show a minimum in their resistivities 
at $T_\rho=50-100$ K and by applying pressure $T_\rho$ increases,\cite{Mori} 
except in $\theta$-(ET)$_2$I$_3$ exhibiting superconducting behavior 
below 3.6 K.\cite{Hayao} 
Their magnetic susceptibilities show Curie-like behavior with no sign of 
magnetic order down to low temperatures. \cite{Nakamura,Nakamura2} 
On the other hand, 
in the $\theta_d$-phase located in the right side of the phase diagram
($\phi \gsim 110^{\circ}$),
the resistivity shows an insulating behavior ($d\rho/dT<0$)\cite{Mori,Komatsu} 
below room temperature 
and its magnetic susceptibility follows that of the 1D 
Heisenberg model ($J \sim 160$ K for $\theta_d$-(ET)$_2$RbZn(SCN)$_4$) 
but exhibits a spin gap behavior at low temperature
in some compounds.\cite{Mori,Komatsu,Takhirov} 
This spin gap behavior is also confirmed by 
$^{13}$C-NMR measurements.\cite{Miyagawa,Nakamura2} 
The CO was recently found 
in the $\theta_d$-phase of (ET)$_2$RbZn(SCN)$_4$\cite{Miyagawa,Chiba}
which is located in the critical region between 
the two phases.\cite{Nakamura2} 
At room temperature it has $\theta$-type structure, 
and upon cooling the sample slowly, 
the resistivity suddenly increases at $T_{\rm str}=195$ K
where it is believed that a structural phase transition takes place 
from $\theta$ to $\theta_d$-type structure,
while this structural phase transition is absent 
if the sample is cooled rapidly so that 
the $\theta$-type structure is maintained down to low temperatures. 
In the slowly cooled case, the $^{13}$C-NMR measurements  
show the existence of the charge disproportionation 
among ET molecules in the $\theta_d$-phase below $T_{\rm str}$, 
i.e. the occurence of a CO state.\cite{Miyagawa,Chiba} 

As for theoretical studies on the CO phenomena 
in 1/4-filled organic conductors, 
most are limited to those in one-dimensional (1D) cases. 
The effects of {\it intersite} Coulomb interactions 
$V_{i,j}$ in such quasi 1D systems 
have been studied from early days\cite{1DextHM} 
and it is known that $V_{i,j}$ 
makes the system toward the 4$k_{\rm F}$ CDW, 
i.e. CO,
which is actually observed experimentally in (DI-DCNQI)$_2$Ag.\cite{Hiraki} 
In the case of (TMTTF)$_2X$ 
the interplay between this $V_{i,j}$ and 
the on-site Coulomb energy $U$ together with the intrinsic dimerization 
give rise to a intermediated state between the 
dimeric AF state and the CO state,\cite{Seo1D} 
and in the cases of $X=$Br and SCN experiments suggest 
the existence of the expected charge disproportionation.\cite{NakamuraTM,Nad}
As regards the 2D case, 
the importance of this $V_{i,j}$ has been clarified recently 
in a 1/4-filled transition metal oxide NaV$_2$O$_5$, 
where 
a similar interplay between $U$, $V_{i,j}$ and the dimerization is present, 
and its ground state is proposed 
to be a peculiar type of CO state due to this $V_{i,j}$.\cite{SeoNa}  
As for the organic systems, 
the calculations without $V_{i,j}$ for 
$\theta_d$-type structure predict the ground state to be 
Mott insulating state,\cite{Seotheta} 
inconsistent with the existence of CO 
disclosed by the experiments mentioned above. 
Thus, studies including $V_{i,j}$ are needed 
to understand CO phenomena in the above 2D organic conductors, 
although CO itself has already been suggested in $\alpha$-type structures 
from the Hartree-Fock calculations by KF.\cite{Kinoalpha} 

The actual values of this $V_{i,j}$ in ET compounds 
are expected to be relatively large 
due to the fact that the wave function of the MO in molecular conductors 
is rather extended, 
so $U$ is expected to be supressed 
so that to be comparable to neighboring $V_{i,j}$. 
Furthermore, the screening of $V_{i,j}$ cannot be expected in ET compounds 
for the HOMO $\pi$-band which is seperated from the other bands, 
as in the transition metals where the $s$-band screens 
the long range Coulomb interactions between the $d$-electrons. 
Actually several calculations for organic compounds 
estimate the nearest-neighbor Coulomb energy $V$ 
to be $20-50\%$ of $U$.\cite{Ducasse,Mila} 
In this paper we consider the value of $U$ on ET molecule 
to be approximately 0.7 eV, 
which is derived from the analysis of NMR data 
of $\kappa$-(ET)$_2$Cu[N(CN)$_2$]Br 
based on random phase approximation,\cite{Mayaffre}
and the values of the neighboring $V_{i,j}$ to be $0.15-0.35$ eV, 
adopting the ratio $V_{i,j}/U = 0.2-0.5$ 
from the above calculations. 

In this paper, we investigate the effect of 
both the on-site and the intersite Coulomb interactions 
on the electronic states in (ET)$_2X$  
and our goal is not only to understand the electronic properties of 
the $\theta$, $\theta_d$ and $\alpha$-phases but also 
to provide a unified view on the variety of ground states 
of this family. 
In \S\ref{model&form} 
the interrelationship between 
the $\theta$, $\theta_d$, $\alpha$ and $\kappa$-type structures 
based on a simplified model 
is explained 
together with the formulation of the MF calculations 
based on the work of KF,\cite{Kinopaper} 
and also an explanation of the CO states of stripe-type, 
which is an analogy to the phenomena 
first found in Nickel oxides,\cite{Tranquada} is given. 
The results of calculations for each type of structure, 
i.e. $\theta$, $\theta_d$ and $\alpha$-type, are shown in \S\ref{results}, 
together with the results for the simplified model. 
The relevance to the actual compounds is discussed in \S\ref{discussions} 
by comparing them with the experiments and 
a unified view on the electronic properties of (ET)$_2X$ is given. 
\S\ref{summary} is devoted to the summary and the conclusion. 
In the following sections 
we call the $\alpha$-type structures with transfer integrals 
corresponding to $\alpha$-(ET)$_2$I$_3$ 
and $\alpha$-(ET)$_2M$Hg(SCN)$_4$, 
as $\alpha$I$_3$-type and $\alpha M$Hg-type, respectively, for abbreviation. 

\section{Model and Formulation of Mean Field Calculations}\label{model&form}

\subsection{Interrelationship among different polytypes of (ET)$_2X$}
\begin{figure}[b]
\begin{center}
\leavevmode\epsfysize=3.3cm
\epsfbox{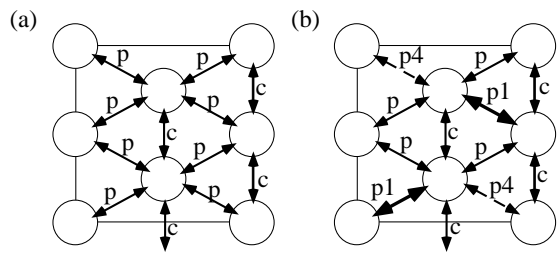}
\end{center}
\caption{A simplified model for (ET)$_2X$ 
introduced by KF.\cite{Kinopaper}
The circles and the arrows represent the ET molecules and the 
intermolecular transfer integrals, respectively. }
\label{fictitious}
\end{figure}
\begin{figure}
\begin{center}
\leavevmode\epsfysize=14cm
\epsfbox{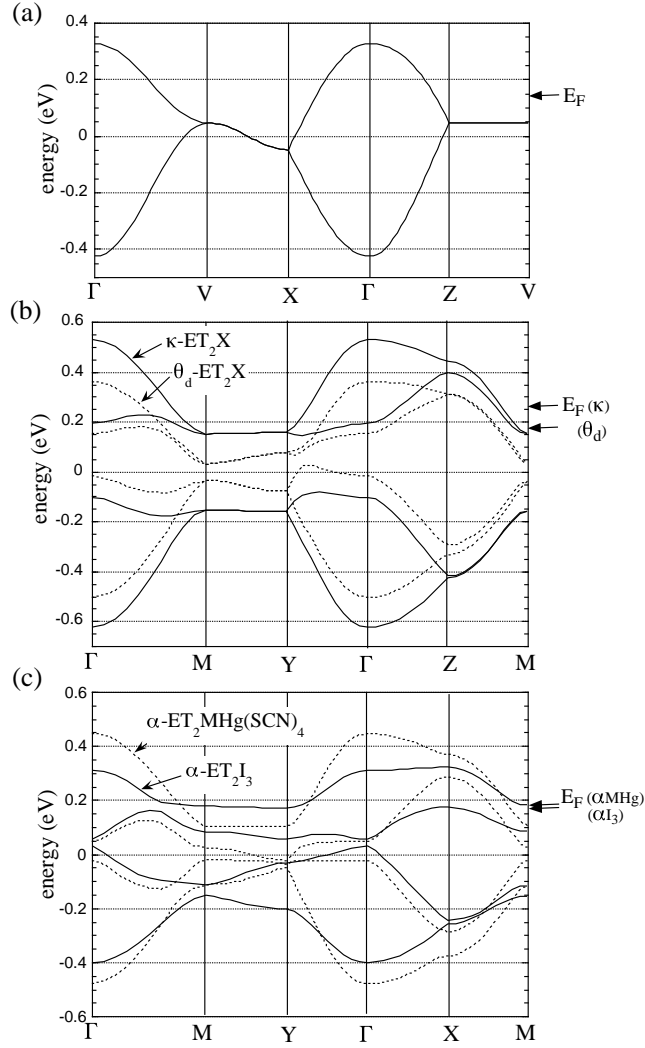}
\end{center}
\caption{The tight binding band structures 
of $\theta$-type (a), $\theta_d$-type and $\kappa$-type (b), 
$\alpha$I$_3$-type and $\alpha M$Hg-type\cite{note}(c) 
structures, 
where the transfer integrals are taken from those 
in Fig. \ref{structure}. 
The fermi energy for each case is also shown.
}
\label{band1}
\end{figure}
\begin{figure}
\begin{center}
\leavevmode\epsfysize=9cm
\epsfbox{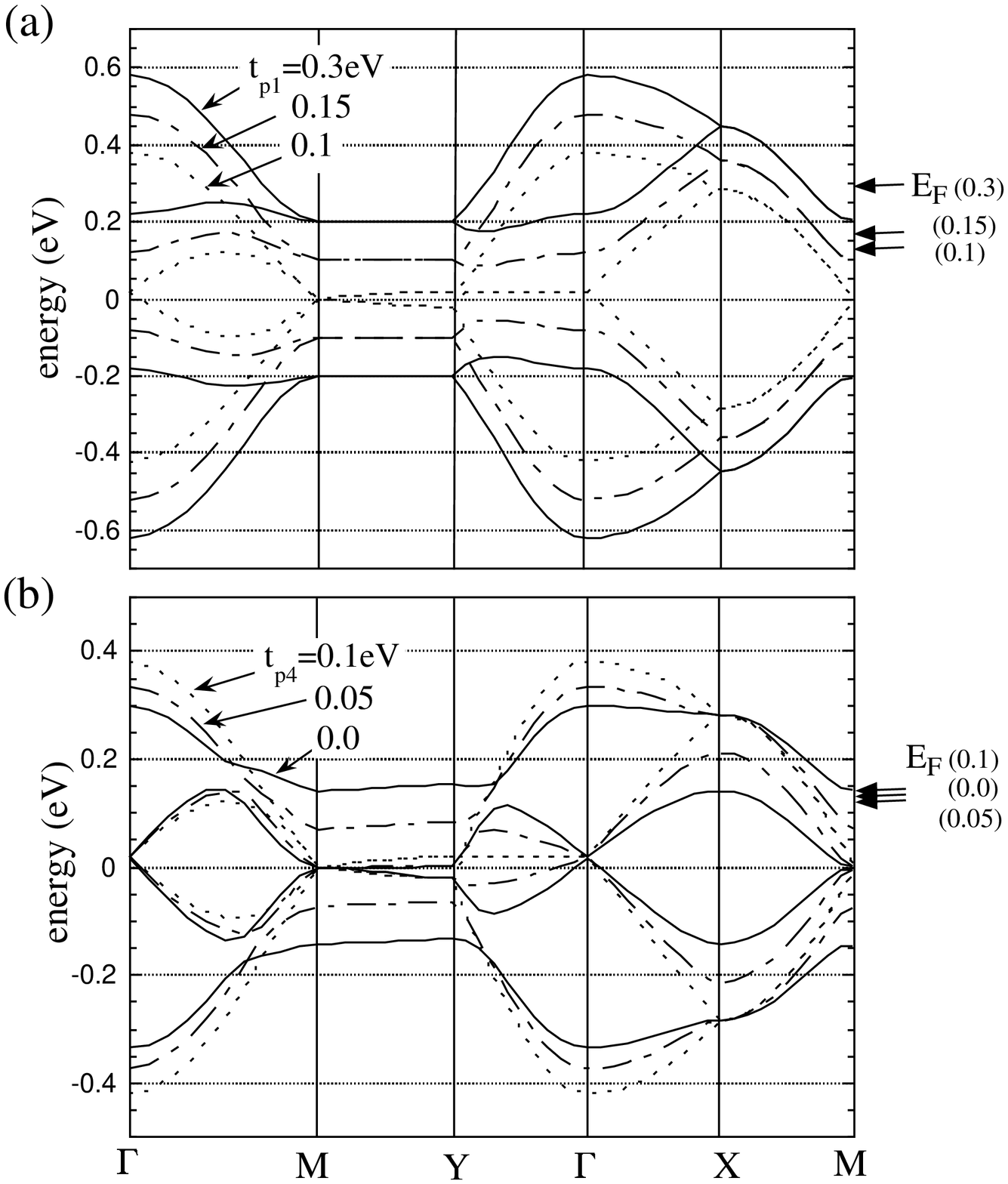}
\end{center}
\caption{The band structures 
for the simplified model shown in Fig. \ref{fictitious} (b). 
$t_c$ and $t_p$ are fixed at $0.01$ eV and $0.1$ eV, respectively. 
The band structures for the case of $t_{p4}=t_p$ 
varying $t_{p1} = 0.1, 0.15, 0.20$ eV (a) 
and for the case of $t_{p1}=t_p$ and $t_{p4} = 0.1, 0.05, 0.0$ eV (b) 
are shown
together with the fermi energy for each case. }
\label{band2}
\end{figure}
It is useful to compare the structural features and the band structures
of different polytypes, 
which has been carried out by KF 
\cite{Kinointer,Kinopaper}
who pursued the interrelationship between 
$\kappa$, $\alpha$I$_3$ and $\alpha M$Hg-type 
structures on the plane of two key ingredients, i.e.  
the degree of dimerization and that of band overlap. 
Here we review their framework  
and also find out where the $\theta$ and $\theta_d$-type 
structures are located, 
by referring a simplified model described in Fig. \ref{fictitious}. 
This model has been introduced by KF 
in order to unify their results 
for the $\alpha$-type and $\kappa$-type structures,
yet they have not performed actual calculation 
on this simplified model. 

In Fig. \ref{band1}, 
the band structures 
of $\theta$-type (a), 
$\theta_d$ and $\kappa$-type (b), 
and $\alpha$I$_3$ and $\alpha M$Hg-type\cite{noteMHg} (c) are shown, 
respectively. 
It can be seen in Fig. \ref{band1}(a) 
that the energy band of $\theta$-type compounds is simply 3/4-filled, 
while in Fig. \ref{band1}(b) for $\theta_d$ and $\kappa$-types
the upper two bands and the lower two bands 
are split due to the dimerization.  
Especially, the band of $\kappa$-phase has a clear dimerization gap 
while the splitting is not appreciable in the $\theta_d$-phase, 
since the relative value of the interdimer transfer integral 
compared to other transfer integrals, 
i.e. the degree of dimerization, 
is quite larger in the $\kappa$-type than in the $\theta_d$-type; 
the interdimer transfer integral in the $\kappa$-type structure 
is $0.257$ eV ($t_{b1}$ in Fig. \ref{Oshima}) 
which is twice larger than the other transfer integrals 
while the one for the $\theta_d$-type structure 
is $0.144$ eV ($t_{p3}$ in Fig. \ref{structure}(c)), 
slightly larger than the other transfer integrals 
in the transeverse direction, 
$t_{p1}$, $t_{p2}$ and $t_{p4}$. 
The difference among the two $\alpha$-type compounds 
has been clarified in the work of KF, \cite{Kinointer,Kinopaper} 
who indicated that this is the degree of the band overlap 
resulting from the difference in the absolute value 
of the transfer integral $t_{p4}$ 
as shown in Fig. \ref{structure}(a). 
The relatively small value of $|t_{p4}|$ in the $\alpha$I$_3$-type structure 
results in the small band overlap, 
while the energy bands for the $\alpha M$Hg-type structure 
has larger overlap due to the larger value of $|t_{p4}|$, 
as seen in Fig. \ref{band1}(c). 

\begin{figure}
\begin{center}
\leavevmode\epsfysize=4.5cm
\epsfbox{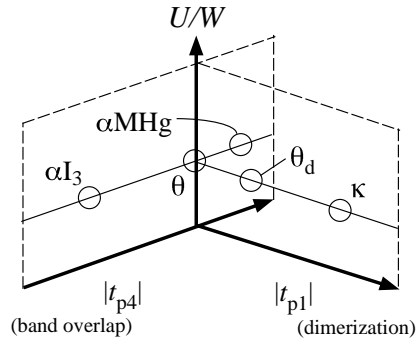}
\end{center}
\caption{Interrelationship between different types of structures 
for (ET)$_2X$, 
schematically shown in the plane of $t_{p1}$ and $t_{p4}$, 
which are the transfer integrals in the model structure 
described in the inset (identical to the one in Fig. \ref{fictitious}(b) ), 
which controls the degree of dimerization and band overlap, respectively. }
\label{relation}
\end{figure}

It turns out that the simplified model in Fig. \ref{fictitious} 
is also an effective model 
for these $\kappa$ and $\theta_d$-type structures by varying $t_{p1}$, 
and for $\alpha$-type structures by varying $t_{p4}$, 
as can be seen in the following. 
The model in Fig. \ref{fictitious}(a) is the $\theta$-type structure 
which has the inversion as well as the spiral symmetry. 
If the absolute value of $t_{p1}$ in Fig. \ref{fictitious}(b) is increased, 
the inversion symmetry is lost, 
and only the spiral symmetry is retained, 
which corresponds to the situation of $\kappa$ and $\theta_d$-type structures. 
If $t_{p4}$ in Fig. \ref{fictitious}(b) is varied, 
then only the inversion symmetry is retained, 
which is the situation of $\alpha$-type structures. 
In Fig. \ref{band2}, the band structures 
for the model in Fig. \ref{fictitious}(b) are shown 
for different parameters, which are $t_a = 0.01$ eV, 
$t_{b4} = t_b = 0.1$ eV and $t_{b1} = 0.1,0.2,0.3$ eV (a),   
and $t_a = 0.01$ eV, 
$t_{b1} = t_b = 0.1$ eV and $t_{b4} = 0.1,0.05,0.0$ eV (b).  
The general features seen in Figs. \ref{band1} (b) and (c)  
are well reproduced in Figs. \ref{band2} (a) and (b), respectively.  
Thus the calculations on this simplified model will 
provide us the idea of a unified view of the (ET)$_2X$ family. 
The interrelationship between the phases can be 
summarized in Fig. \ref{relation}, 
which is similar to the corresponding figure in ref. \ref{Kinointer} 
but the locations of the $\theta$ and $\theta_d$-phases beeing added. 

\subsection{Formulation}

2D layer of ET molecules is considered 
by taking into account the transfer integrals betweeen them 
and the mutual Coulomb interaction among electrons. 
In this work the treatment of KF is basicly followed,\cite{Kinopaper}
but here the intersite Coulomb interactions are also included.  
Thus our Hamiltonian is an extended Hubbard one, 
which is as follows; 
\begin{eqnarray}
H=\sum_{<i,j>}&\sum_{\sigma}&\left(t_{i,j}a^{\dagger}_{i\sigma}
    a_{j\sigma}+h.c.\right) \nonumber\\
   &+&\sum_{i} U n_{i\uparrow}n_{i\downarrow}
    +\sum_{<i,j>}V_{i,j} n_in_j,
\label{eqn:Hamil}
\end{eqnarray}
where $<i,j>$ denotes the neighboring site pair, 
$\sigma$ is the spin index which takes $\uparrow$ and $\downarrow$, 
$n_{i\sigma}$ and  $a^{\dagger}_{i\sigma}$ ($a_{i\sigma}$) denote 
the number operator and the creation (annihilation) operator for the 
electron of spin $\sigma$ at the $i$th site, respectively, 
and $n_i=n_{i\uparrow}+n_{i\downarrow}$.  

The Coulomb interactions $U$, $V_{i,j}$ are treated 
within the MF approximation 
in a manner similar to that in refs. 
\ref{Kinoalpha}-\ref{Kinopaper},\ref{Seo1D},\ref{SeoNa}; 
\begin{eqnarray}
n_{i\uparrow}n_{i\downarrow} &\rightarrow& 
\left<n_{i\uparrow}\right>n_{i\downarrow}
+n_{i\uparrow}\left<n_{i\downarrow}\right>
-\left<n_{i\uparrow}\right>\left<n_{i\downarrow}\right> \nonumber\\
n_in_j &\rightarrow&  \left<n_i\right>n_j
+n_i\left<n_j\right>
- \left<n_i\right>\left<n_j\right>, 
\end{eqnarray}
which corresponds to the Hartree approximation. 
The MF Hamiltonian in $k$-space is given by
\begin{full} 
\begin{eqnarray}
H^{\rm MF}=\sum_{k\sigma}\left(\begin{array}{cccc}
a_{1k\sigma}\\
a_{2k\sigma}\\
\vdots\\
a_{mk\sigma} \end{array}\right)^{\dagger}
\left[h_{0}+U\left(\begin{array}{cc}
\begin{array}{cc}n_{1\bar{\sigma}} & \\
 & n_{2\bar{\sigma}}
\end{array} & 0 \\
0 & \begin{array}{cc}\ddots & \\
 & n_{m\bar{\sigma}}
\end{array}
\end{array}\right) + h_V \right]
\left(\begin{array}{cccc}
a_{1k\sigma}\\
a_{2k\sigma}\\
\vdots\\
a_{mk\sigma} \end{array}\right),
\label{eqn:HFHamil}
\end{eqnarray}
\end{full}
where $h_0$ and $h_V$ denote the matrices for the 
kinetic energy term and the intersite Coulomb interaction term, respectively.
The actual form of these matrices depends on each structure 
and unit cell size 
whose examples are shown in \S\ref{stripeCO}. 

The crysatallographic unit cell in the layer 
for the $\theta$-type structure 
is approximated as rectangle with $a=2c$ 
and those for the $\theta_d$ and $\alpha$-type structures 
as square. 

Self-consistent solutions are searched for at $T=0$ 
with the average electron density being fixed at 1.5 per ET site 
by considering several unit cell sizes 
and their energies are compared
so that the ground state can be decided.
The solutions where the net magnetic moment of the whole system is finite, 
i.e. ferro/ferrimagnetic solutions are neglected, 
as in the work of KF, 
since the stability of these states, if it occurs, 
can be thought to be the artificat of the approximation. 
The total energy ${\cal E}$ is calculated as 
\begin{eqnarray}
{\cal E}=\frac 1{N_{\rm cell}}&\sum_{lk\sigma}&\epsilon_{lk\sigma}
n_{\rm F}\left(\epsilon_{lk\sigma}\right) \nonumber\\
-\sum_i&U&\left<n_{i\uparrow}\right>\left<n_{i\downarrow}\right>
-\sum_{<i,j>}V_{i,j} \left<n_i\right>\left<n_j\right>,
\end{eqnarray}
where $N_{\rm cell}$ is the total number of the cells, 
$\epsilon_{lk\sigma}$ is the $l$th eigenvalue of the 
$m\times m$ Hamiltonian matrix in eq. \ref{eqn:HFHamil} 
at each $k$ and 
$n_{\rm F}\left(\epsilon \right)$ is the Fermi distribution function. 

The intersite Coulomb interactions we consider are taken 
along the bonds with transfer integrals in Fig. \ref{structure}, 
i.e. those between the neighboring ETs. 
For simplification only two kinds of values are considered:  
the intersite Coulomb energy along the stacking direction 
$V_c$ which is the one between parallel molecules, 
and along the bonds in the transverse directions $V_p$
which is the one between molecules with dihedral angle $\phi$. 
In the actual compounds, 
the values of intermolecular Coulomb energy should be different 
for the bonds with different transfer integral 
since the configurations between pair of molecules are different. 
So the above model with only $V_c$ and $V_p$ 
should be appropriate for the $\theta$-type structure 
with only two kinds of values of transfer integrals $t_c$ and $t_p$. 
However we approximate 
in the calculations also for the $\theta_d$ and $\alpha$-type structures 
that only two kinds of the intersite Coulomb energy $V_c$ and $V_p$ exist 
since the deviation in the actual $\theta_d$ and $\alpha$-(ET)$_2X$, 
which both can be considered as structurally similar phases 
to the $\theta$-phase,\cite{TMoriGenealogy} 
is expected to be small. 

\subsection{Stripe-type charge ordered state}\label{stripeCO}
\begin{figure}
\begin{center}
\leavevmode\epsfysize=8.3cm
\epsfbox{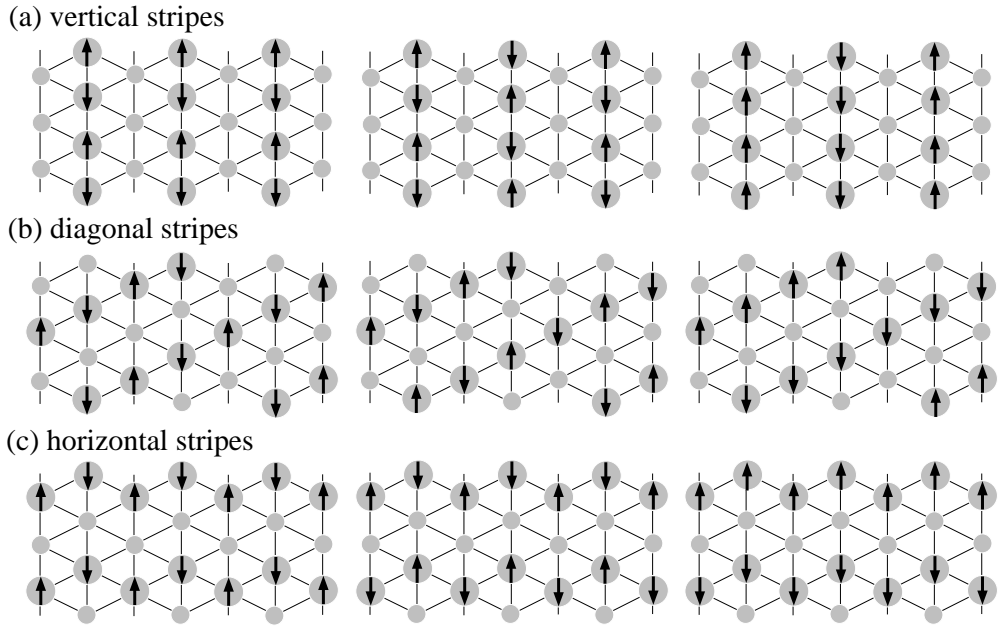}
\end{center}
\caption{\protect\vspace*{-2.5mm}\protect\\
\hspace*{-0.2cm}
\parbox{17cm}{\hspace*{1cm}
Schematic representation of the spin moments and hole densities 
for the stripe-type CO solutions found to be stabilized. 
The diameters of the gray circles and the arrows 
represent the hole densities $2-<n_i>$ and the spin moments, 
respectively.
}}
\label{stripe}
\end{figure}

\begin{figure}
\begin{center}
\leavevmode\epsfysize=5.8cm
\epsfbox{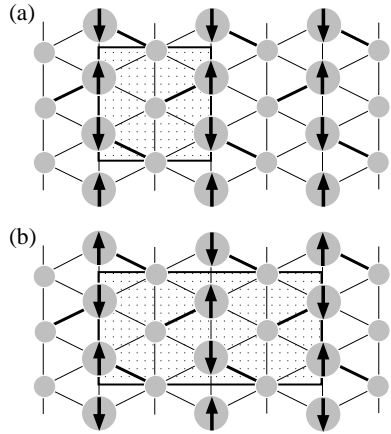}
\end{center}
\caption{The magnetic cell for two examples of  
vertical stripe-type CO solution for $\theta_d$-type structure. 
Thick lines represent the bonds with the largest transfer integral $t_{p3}$. 
The dotted areas, the larger gray circles and the arrows denote 
the unit cells, the sites with excess holes and the magnetic moments, 
respectively.}
\label{example}
\end{figure}

In the following sections the solutions with charge disproportionation 
of stripe-type are found to be relevant in some compounds, 
where the sites with more amount of charge  
are arranged in rows alternately. 
Three kinds of charge patterns, namely arrangements of holes, 
can be naturally considered as shown in Fig. \ref{stripe}, 
which we call 
the vertical stripes (a), the horizontal stripes (b) 
and the diagonal stripes (c), respectively. 
In Fig. \ref{stripe}, 
the schematic representation of the solutions 
with stripe-type CO which are found to be stabilized 
are shown. 
In Fig. \ref{example}(a) and (b) 
two examples for the CO state with vertical stripes 
for the $\theta_d$-type structure are displayed 
where the unit cells consist of 4 and 8 ETs
($m=4$ and $m=8$ in eq. \ref{eqn:HFHamil}), respectively, 
where the matrix elements of $h_0$ and $h_V$ 
for the former case of Fig. \ref{example}(a) are written as 
%
%
%
\begin{center}
\begin{eqnarray}
\left[h_0\right]_{ii} &=& 0, \nonumber\\
\left[h_0\right]_{12} &=& \bigl(\left[h_0\right]_{34}\big)^* =
t_{c1}e^{ik_y/2}+t_{c2}e^{-ik_y/2}, \nonumber\\ 
\left[h_0\right]_{13} &=& 2t_{p2}e^{-ik_y/4}cos(k_x/2), \nonumber\\
\left[h_0\right]_{14} &=& t_{p3}e^{i(k_x/2+k_y/4)}
+t_{p1}e^{-i(k_x/2-k_y/4)}, \nonumber\\
\left[h_0\right]_{23} &=&  t_{p1}e^{i(k_x/2+k_y/4)}
+t_{p3}e^{-i(k_x/2-k_y/4)}, \nonumber\\
\left[h_0\right]_{24} &=& 2t_{p4}e^{-ik_y/4}cos(k_x/2), \nonumber\\
\left[h_0\right]_{ij} &=& \bigl(\left[h_0\right]_{ji}\bigr)^* 
\hspace{0.3cm}(i>j), \nonumber
\label{eqn:h0}
\end{eqnarray}
\end{center}
\begin{center}
\begin{eqnarray}
\left[h_V\right]_{11} &=& 2V_cn_2+2V_p(n_3 +n_4), \nonumber\\
\left[h_V\right]_{22} &=& 2V_cn_1+2V_p(n_3 +n_4), \nonumber\\
\left[h_V\right]_{33} &=& 2V_cn_4+2V_p(n_1 +n_2), \nonumber\\
\left[h_V\right]_{44} &=& 2V_cn_3+2V_p(n_1 +n_2), \nonumber\\
\left[h_V\right]_{ij} &=& 0 \hspace{0.3cm} (i\neq j).\nonumber
\label{eqn:hV}
\end{eqnarray}
\end{center}
%

\section{Results of Calculations}\label{results}
In self-consistent MF calculations, 
the 2D integrations are approximated by a sum of the mesh, 
where the analytical tetrahedron method is employed\cite{Kinopaper,ATM}
for 30$\times$30 $k$-points 
per Broullian zone for the cell of four sites. 
In \S \ref{theta} - \ref{alpha}, 
the value of $U$ is fixed at $U=0.7$ eV. 
The dependences on $U$ are considered in \S\ref{model}. 

\subsection{$\theta$-type}\label{theta}
The results for $\theta$-type structure reveal that 
stripe-type CO states have the lowest energy 
for relevant values of Coulomb energies, 
though the charge pattern depends on the parameters of the model, 
which are the transfer integrals $t_c$ and $t_p$ and 
the intersite Coulomb energies $V_c$ and $V_p$. 
First we consider the case of $t_c = 0.01$ eV and $t_p = -0.1$ eV, 
the typical values for 
the series of $\theta$-(ET)$_2MM'$(SCN)$_4$,\cite{Mori} 
and the case of the isotropic intersite Coulomb interactions, 
i.e. $V_c=V_p \equiv V$. 
\begin{figure}
\begin{center}
\leavevmode\epsfysize=10cm
\epsfbox{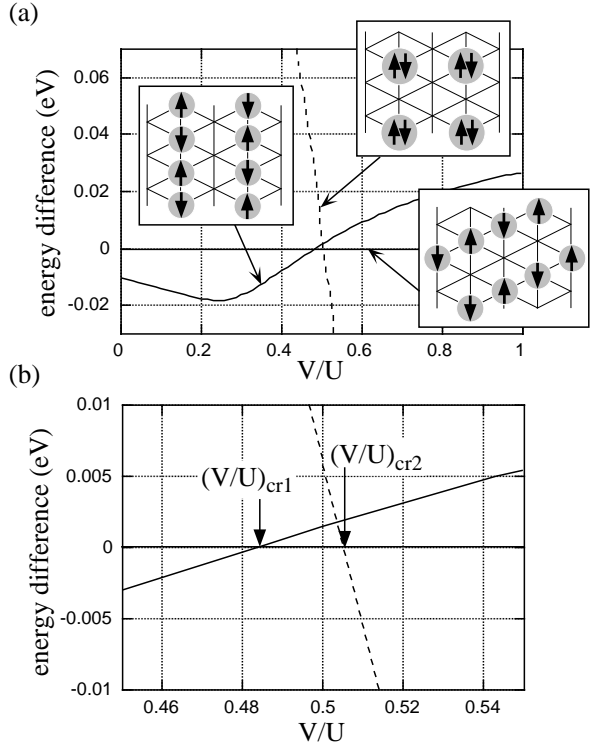}
\end{center}
\caption{$V/U$ dependence of relative energy 
compared to the diagonal stripe solution 
which is schematically shown in an inset, 
for $U = 0.7$ eV, $t_c=0.01$ eV and $t_p=-0.1$ eV.  
The schematic representations of the charge pattern 
as well as the magnetic structure in these solutions are shown in insets.
The enlargement of the critical region is shown in (b). 
}
\label{en-V_th_U07}
\end{figure}
In Fig. \ref{en-V_th_U07}, we show the $V$-dependence 
of the calculated relative energies for the solutions 
which are stable for $0\leq V/U \leq 1$ 
compared to the energy of the diagonal stripe solution 
schematically shown in an inset.  
The solutions which are not shown in the figure
have higher energies. 
It can be seen that the vertical stripe solution with AF ordering 
both along and between these stripes (see inset of Fig. \ref{en-V_th_U07}), 
has the lowest energy for $V/U\leq (V/U)_{\rm cr1} = 0.49$, 
whereas for $(V/U)_{\rm cr1} \leq V/U \leq (V/U)_{\rm cr2}=0.51$, 
the state with diagonal stripes does. 
For $V/U \geq (V/U)_{\rm cr2}$ 
a nonmagnetic insulating state is stable, 
which can be seen as a bipolaronic state 
with every fourth site being occupied by almost two holes, 
as schematically shown in an inset. 
\begin{figure}[t]
\begin{center}
\leavevmode\epsfysize=7cm
\epsfbox{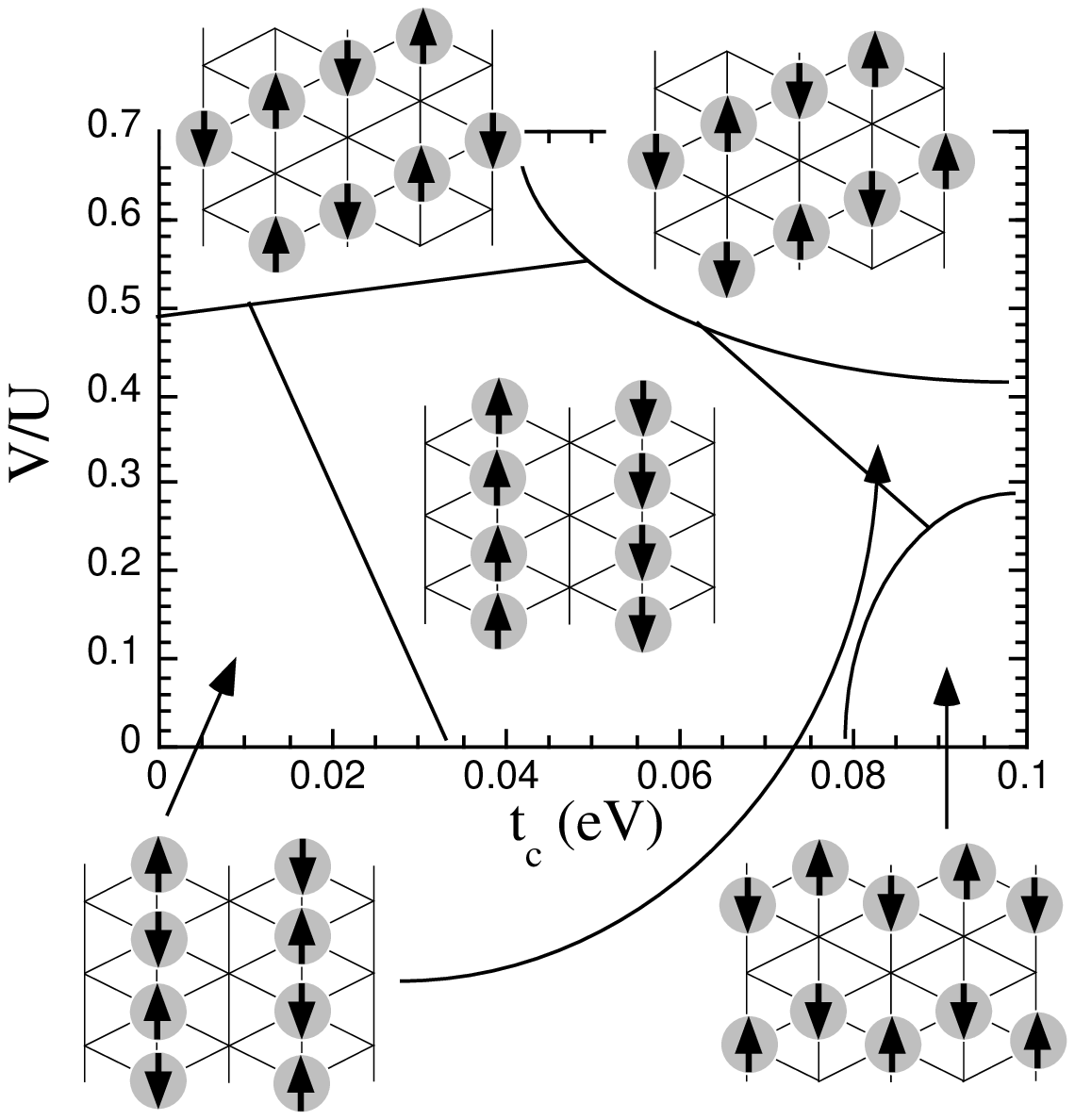}
\end{center}
\caption{The phase diagram for the $\theta$-type structure 
on the plane of $V/U$ and $t_c$, for $U=0.7$ eV and $t_p=-0.1$ eV, 
excluding the bipolaronic state. 
}
\label{phasetheta}
\end{figure}

The value of $(V/U)_{\rm cr1}$ slightly varies 
when the values of transfer energies, $t_c$ and $t_p$, are varied, 
though the qualitative behaviors are similar 
for $t_c/|t_p| \lsim 0.2 $, which is appropriate for 
$\theta$-(ET)$_2MM'$(SCN)$_4$.\cite{Mori} 
In Fig. \ref{phasetheta} the phase diagram for the $\theta$-type structure 
is shown on the plane of $V/U$ and $t_c$, 
with fixed values of $U=0.7$ eV and $t_p=-0.1$ eV, 
excluding the bipolaronic state 
to see the competition between different stripe-type CO states. 
It can be seen that, concerning the charge pattern, 
the vertical stripes are preferred in the region of small $V$
while the diagonal stripes are in the region of large $V$
as an overall feature, 
though the spin configurations depend on the value of $t_c$. 
\begin{figure}[t]
\begin{center}
\leavevmode\epsfysize=5cm
\epsfbox{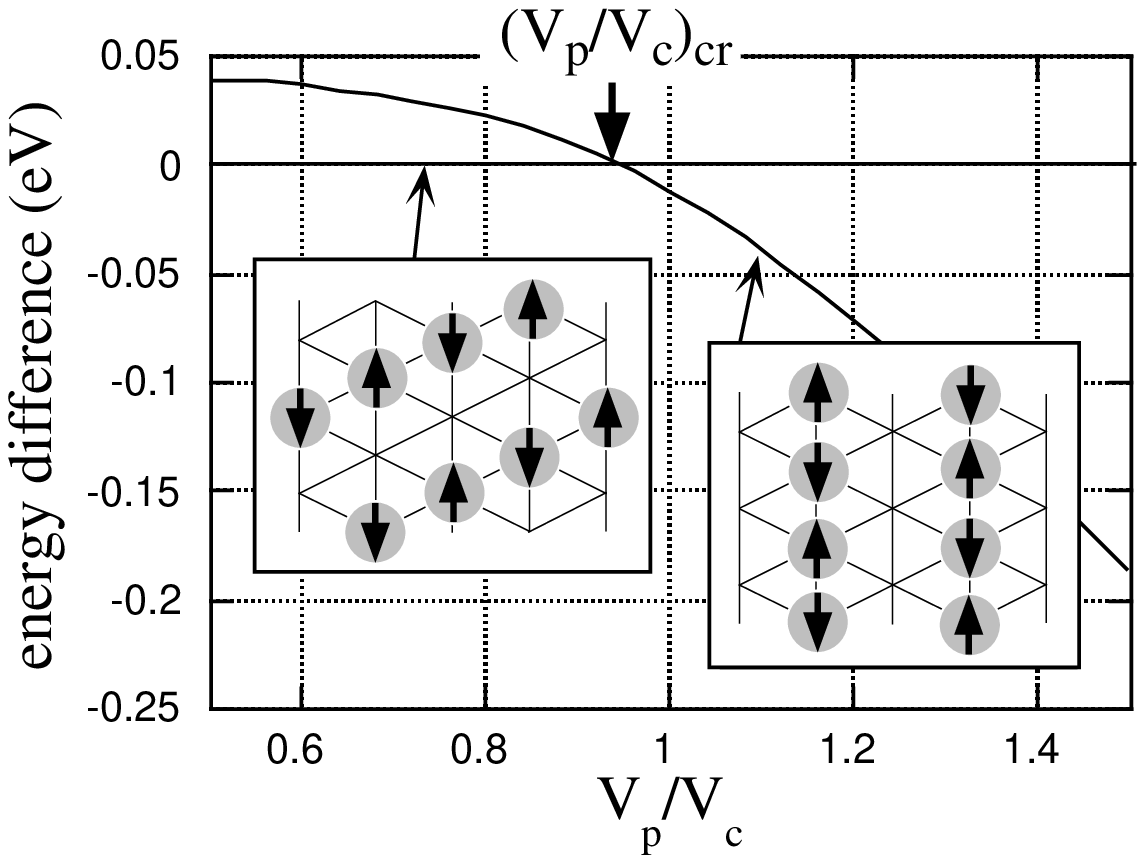}
\end{center}
\caption{$V_p/V_c$ dependence of relative energy 
compared to the diagonal stripe solution 
which is schematically shown in the right inset, 
on the $\theta$-type structure for the fixed value of $V_c=0.25$ eV 
and $U=0.7$ eV.}
\label{en-Vp_th_U07V025}
\end{figure}

Next we vary the value of $V_p/V_c$ 
yet fixing $U=0.7$ eV, $t_c=0.01$ eV and $t_p=-0.1$ eV. 
In Fig. \ref{en-Vp_th_U07V025}, 
the $V_p/V_c$-dependence of the relative energy
of the vertical stripe solution compared to the diagonal stripe solution 
for the case of $V_c$ fixed at 0.25 eV is shown. 
Note that the other solutions have higher energies than these two solutions. 
As $V_p/V_c$ is decreased, there exists a phase transition 
at $(V_p/V_c)_{\rm cr}=0.95$ 
from the CO state with vertical stripes to 
the one with diagonal stripes. 
It is noteworthy that a very slight deviation from $V_p/V_c=1$, 
where the state with vertical stripes is stable, 
makes the energy of the diagonal stripe solution lower, 
so the boundary betweeen the vertical 
and diagonal stripes in Fig. \ref{phasetheta}, 
which is around $V/U =$ 0.5 for the isotropic case, 
is easily varied when $V_p/V_c$ is varied. 

\subsection{$\theta_d$-type} \label{thetad}
\begin{figure}[t]
\begin{center}
\leavevmode\epsfysize=8cm
\epsfbox{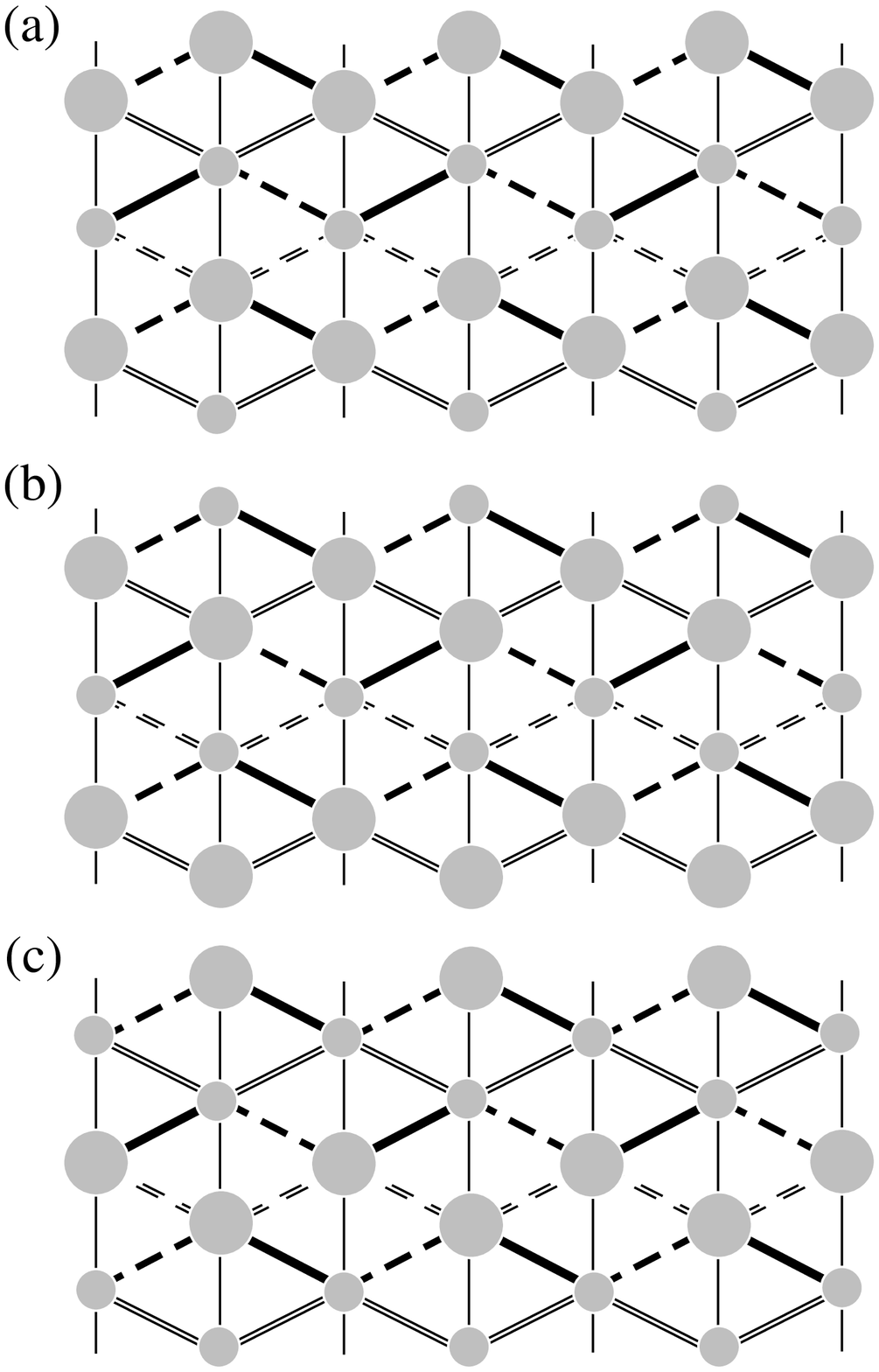}
\end{center}
\caption{Schematic representations of the charge pattern 
the three different horizontal stripes 
in $\theta_d$-type structure:  
stripes along $t_{p3}$ and $t_{p1}$ (a), 
those along $t_{p4}$ (b) and those along $t_{p2}$ (c). 
The thick dotted, double dotted, thick solid, double solid bonds 
represent the bonds with the transfer integral 
$t_{p1}$, $t_{p2}$, $t_{p3}$, $t_{p4}$, 
respectively and the thin solid bonds are for $t_{c1-2}$. 
}
\label{thdyoko}
\end{figure}

\begin{figure}[b]
\begin{center}
\leavevmode\epsfysize=3cm
\epsfbox{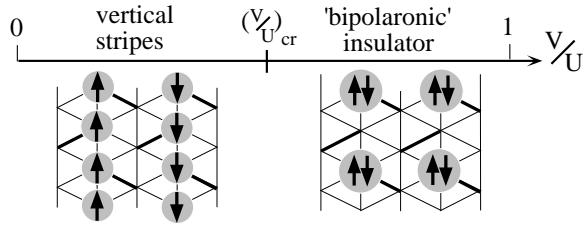}
\end{center}
\caption{The sequences of the solution which gives the lowest energy for 
$\theta_d$-type structure, as a function of $V/U$ ($U=0.7$ eV). 
The thick bonds represent the bonds with the largest transfer energy $t_{p3}$, 
as in Fig. \ref{example}. }
\label{en-V_thd_U07}
\end{figure}

\begin{figure}[t]
\begin{center}
\leavevmode\epsfysize=9cm
\epsfbox{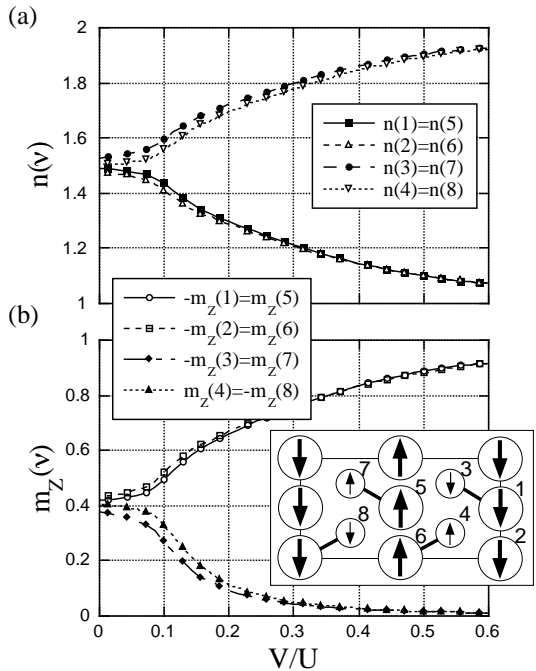}
\end{center}
\caption{$V/U$ dependence of the charge densities (a) 
and magnetic moments (b) 
on each site for the state with vertical stripes which gives the lowest energy 
in the region of small $V$ for the $\theta_d$-type structure.
Inset shows the indices for sites 
and schematically the charge pattern and 
the magnetic structure in this vertical stripe solution,  
where the thick bonds describe the bonds with 
largest transfer energy $t_{p3}$,  
the sites with large (small) arrows and those with large (small) diameters 
represent the sites with large (small) hole density 
and those with large (small) magnetic moment 
while the direction of the arrows shows the direction of spins. }
\label{nSz-V_thd}
\end{figure}

The calculations for the $\theta_d$-type structure
show that several different types of charge patterns compete 
with each other when the values of intersite Coulomb energies are varied.
In this subsection the transfer integrals are fixed 
to those in Fig. \ref{structure}(c). 
In contrast to the $\theta$-type structure above, 
there exist three different charge patterns 
for horizontal stripe solutions 
which have different energies: 
the solutions with excess holes on sites 1 and 3 
(stripes along $t_{p3}$ and $t_{p1}$), 
sites 1 and 4 (stripes along $t_{p4}$), 
and sites 2 and 3 (stripes along $t_{p2}$) 
as shown in Fig. \ref{thdyoko}(a)$\sim$(c), respectively.  
Among these states, 
the stripes along $t_{p3}$ and $t_{p1}$ have lower energies 
in the region of large $V$ ($V/U \gsim 0.5$), 
whereas for small $V$ ($V/U \lsim 0.5$) 
the stripes along $t_{p4}$ does. 
In Fig. \ref{en-V_thd_U07}, 
the ground state solutions are shown 
for the case of isotropic intersite Coulomb interactions, 
$V_c=V_p \equiv V$,  
as $V/U$ is varied with $U$ fixed at 0.7 eV. 
It can be seen there that when $V$ is increased 
the charge pattern of the solution having the lowest energy 
varies from the vertical stripes to the bipolaronic state 
at $V/U = (V/U)_{\rm cr} = 0.50 $. 
In Fig. \ref{nSz-V_thd}, the magnetic moment 
and the charge density on each site 
for the vertical stripe solution 
are shown as a function of $V/U$. 
As $V$ is decreased, this stripe-type CO state continuously crosses over 
to the dimeric AF state, 
with almost same charge density on each site 
and AF ordering between dimers, 
i.e. pairs of molecules connected by the largest transfer energy $t_{p3}$, 
which is known to be stable for $V=0$.\cite{Seotheta}

\begin{figure}
\vspace*{0.5cm}
\begin{center}
\leavevmode\epsfysize=4cm
\epsfbox{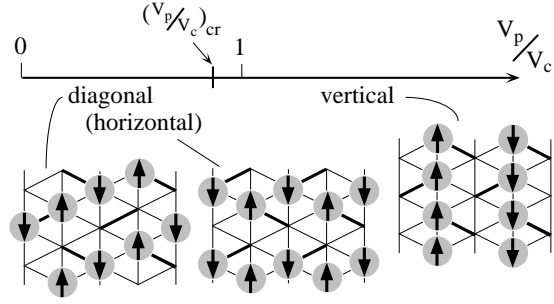}
\end{center}
\caption{The $V_p/V_c$-dependence of the solution which gives the 
lowest energy for $\theta_d$-type structure, 
for the case of $V_c = 0.25$ eV and $U=0.7$ eV. 
}
\label{en-Vp_thd_U07V025}
\end{figure}

When $V_p/V_c$ is varied, the solution which gives the lowest 
energy easily changes as in the case of $\theta$-phase. 
In Fig. \ref{en-Vp_thd_U07V025}, we show the 
variation of ground state solutions for the fixed value of $V_c$=0.25 eV. 
For $V_p/V_c \leq (V_p/V_c)_{\rm cr}= 0.88 $, 
the state with diagonal stripes is stable, 
whereas there exists another solution with almost same energy: 
the horizontal stripes along $t_{p4}$, 
as shown in Fig. \ref{en-Vp_thd_U07V025}. 
For $V_p/V_c \geq (V_p/V_c)_{\rm cr}$ 
the vertical stripe solution has the lowest energy, 
as in the case of $\theta$-phase. 

\subsection{$\alpha$I$_3$-type and $\alpha M$Hg-type} \label{alpha}
\begin{figure}
\begin{center}
\leavevmode\epsfysize=6cm
\epsfbox{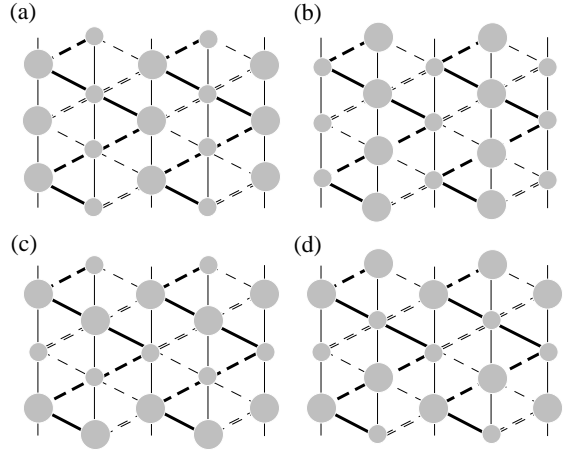}
\end{center}
\caption{Schematic representations of the charge patterns for the 
different horizontal stripes and vertical stripes 
in the $\alpha$-type structure: 
vertical stripes along stack I (a), those along stack II (b), 
horizontal stripes along $t_{p2}$ and $t_{p3}$ (c) 
and those along $t_{p1}$ and $t_{p4}$ (d). 
The thick dotted, thick solid, double dotted, thin dotted bonds 
represent the bonds with transfer integrals 
$t_{p1}$, $t_{p2}$, $t_{p3}$, $t_{p4}$, 
respectively and the thin solid bonds are for $t_{c1 \sim 3}$. 
}
\label{alcharge}
\end{figure}

The calculations for the $\alpha$-type structures show that 
the diagonal stripe solutions have 
rather high energies compared to the other stripe-type CO states, 
which is in contrast to the above cases for 
the $\theta$ and $\theta_d$-type structures. 
In this subsection the transfer integrals are fixed to those 
in Fig. \ref{structure}(a). 
Due to the lower symmetry than the $\theta$-type structure, 
as in the $\theta_d$-phase, 
concerning the vertical and the horizontal stripes 
each has two solution with different charge patterns
having different energies: 
as for the vertical stripes those along stack I and those along stack II, 
and for the horizontal stripes those along $t_{p2}$ and $t_{p3}$ 
and those along $t_{p1}$ and $t_{p4}$,  
which are shown in Fig. \ref{alcharge}.
\begin{figure}
\begin{center}
\leavevmode\epsfysize=9cm
\epsfbox{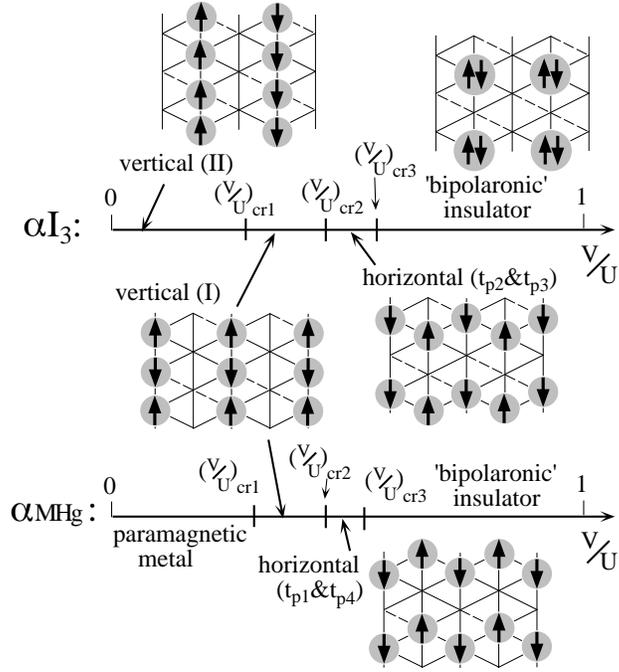}
\end{center}
\caption{The sequences of the solution which gives the lowest energy for 
$\alpha$-type structures, as a function of $V/U$ ($U=0.7$ eV). 
The dotted bonds represent the transfer integral $t_{p4}$. 
The distinction between different stripes among the vertical stripes 
on stack I or II and among the horizontal stripes 
along $t_{p2}$ and $t_{p3}$, 
and $t_{p1}$ and $t_{p4}$ are also described here.}
\label{en-V_al_U07}
\end{figure}
\begin{figure}
\begin{center}
\leavevmode\epsfysize=3.5cm
\epsfbox{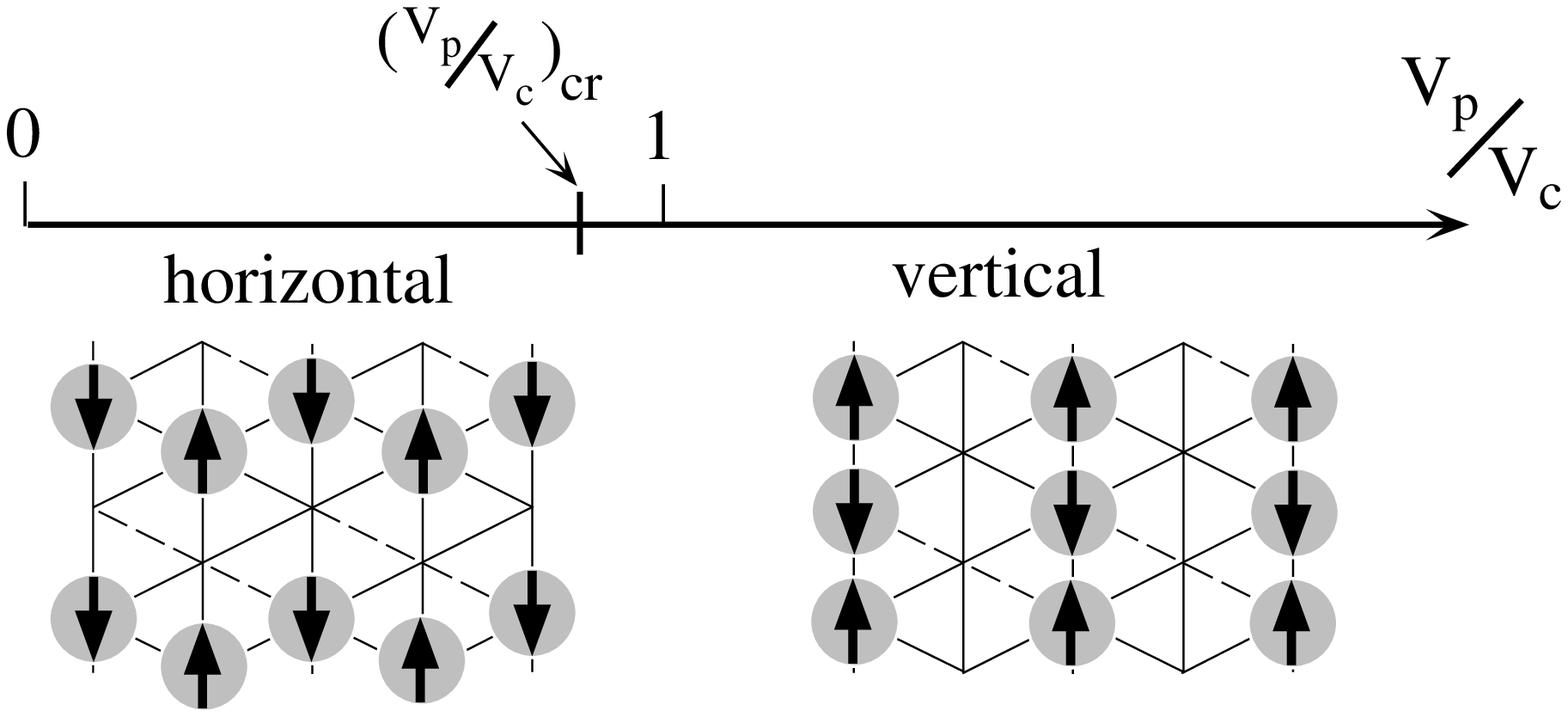}
\end{center}
\caption{$V_p/V_c$ dependence of the solution which gives the lowest energy 
for the $\alpha$I$_3$-type structure with $V_c = 0.25$ eV and $U=0.7$ eV.}
\label{en-Vb_al_U07V025}
\end{figure}

In the upper part of Fig. \ref{en-V_al_U07}, 
the sequences of ground state solutions 
for the $\alpha$I$_3$-type structure 
for the case of isotropic intersite Coulomb interactions, 
$V_c=V_p \equiv V$, are shown. 
As $V$ is increased, the state which is stable 
varies from the one with vertical stripes on stack II, 
to the one with another vertical stripes on stack I 
at $V/U = (V/U)_{\rm cr1} = 0.27$. 
KF concluded that the latter state was the ground state 
of $\alpha$-(ET)$_2$I$_3$ for $V=0$ 
under the assumption that the magnetic cell consists of four molecules 
which is identical to the structural unit cell,\cite{Kinoalpha} 
but the considerations of larger cells in the present calculation 
indicate that at $V=0$ 
the vertical stripe solution with the other charge pattern, 
i.e. the stripes on stack II, has lower energy. 
As $V$ is increased further, 
a phase of horizontal stripes along $t_{p2}$ and $t_{p3}$ 
appears at $V/U = (V/U)_{\rm cr2} = 0.43$ 
and a bipolaronic state takes place 
above $V/U = (V/U)_{\rm cr3} = 0.50$. 

In Fig. \ref{en-Vb_al_U07V025}, we show the $V_p/V_c$-dependence of 
the ground state solutions 
for the case of fixed value of $V_c$=0.25 eV. 
As in the calculations for the $\theta$ and $\theta_d$-type structures, 
the solution which has the lowest energy easily changes 
from that for the case of isotropic interstite Coulomb interactions, 
when the value of $V_p/V_c$ is deviated from 1. 
The solution with horizontal stripes along $t_{p2}$ and $t_{p3}$ 
has the lowest energy for $V_p/V_c \leq (V_p/V_c)_{\rm cr} = 0.96$, 
whereas the one with vertical stripes on stack I is stable 
for $V_p/V_c \geq (V_p/V_c)_{\rm cr}$.
\begin{figure}
\begin{center}
\leavevmode\epsfysize=13.5cm
\epsfbox{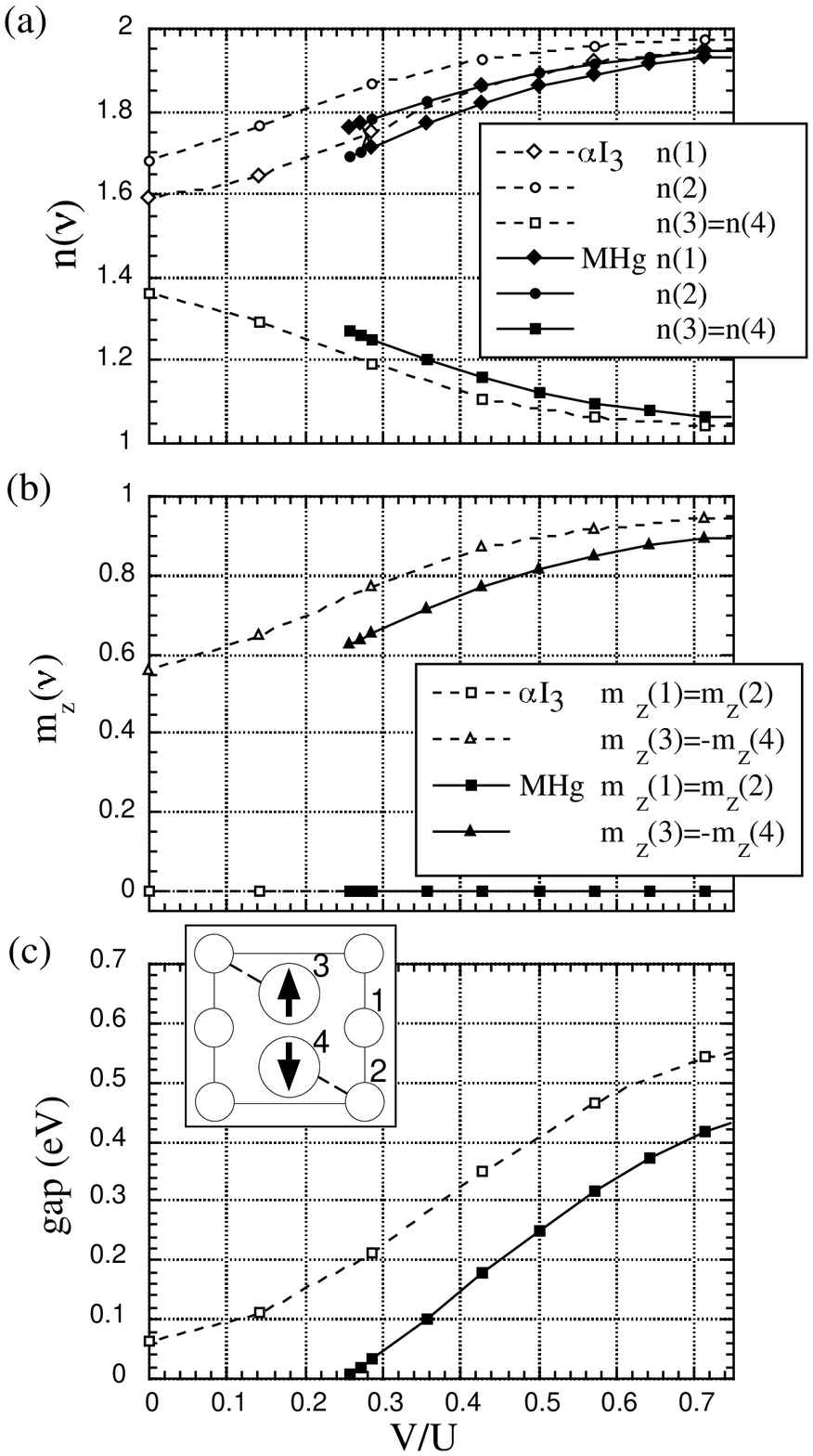}
\end{center}
\caption{$V/U$ dependence of the charge density (a), the magnetic moment (b) 
and the band gap (c) on each site for the solutions 
with vertical stripes on stack I
for $\alpha M$Hg-type  
and $\alpha$I$_3$-type structures for $U=0.7$ eV.}
\label{nSz-V_MHg_U07}
\end{figure}

In the work of KF, Hartree-Fock calculations for $U=0.7$ 
on the $\alpha M$Hg-type structure  
suggest a paramagnetic metallic (PM) ground state.\cite{Kinointer} 
By including $V_{i,j}$ some CO phases appear,  
as can be seen in the lower part of Fig. \ref{en-V_al_U07} 
where the ground state solutions 
for the case of isotropic intersite Coulomb interactions,  
$V_p=V_c\equiv V$, are shown. 
The ground state changes from the PM phase, 
as $V$ is increased, to a CO phase 
with the vertical stripes on stack I at $V/U = (V/U)_{\rm cr1} = 0.32$.  
As $V$ is increased further a phase transition is present 
at $V/U= (V/U)_{\rm cr2} = 0.46$ to a phase of the horizontal stripes 
along $t_{p1}$ and $t_{p4}$, 
and above $V/U= (V/U)_{\rm cr3} =0.54$ 
the bipolaronic state has the lowest energy. 
As can be seen by comparing the upper and lower parts 
of Fig. \ref{en-V_al_U07}, 
the sequences of ground states 
for the $\alpha M$Hg and $\alpha$I$_3$-type structures 
are qualitatively similar 
for $V/U \geq (V/U)_{\rm cr1}$, 
although the charge pattern for the horizontal stripes 
stable for $(V/U)_{\rm cr2} \leq V/U \leq (V/U)_{\rm cr3}$
are different to each other. 
In Fig. \ref{nSz-V_MHg_U07}, 
the charge density, the magnetic moment on each site, and the band gap 
for the solution with vertical stripes on stack I 
for the $\alpha$I$_3$ and $\alpha M$Hg-type structures 
as a function of $V/U$ is plotted 
which are stable for $(V/U)_{\rm cr1} \leq V/U \leq (V/U)_{\rm cr2}$. 
The notation of `band gap' is the same as in the work of KF,\cite{Kinopaper} 
which is the value of the bottom of the 1st band minus the top of the 2nd band 
for the actual case with the unit cell of 4 sites 
($m=4$ in eq. \ref{eqn:HFHamil}). 
It can be seen that both solutions show qualitatively similar features 
though the CO state for  
the $\alpha M$Hg-type structure has relatively small values of band gap 
compared to those for the $\alpha$I$_3$-type structure. 

\subsection{Simplified model for (ET)$_2X$} 
\label{model}
As have been seen in \S \ref{theta} - \ref{alpha}
the CO states are more stabilized 
by the intersite Coulomb interactions $V_{ij}$. 
In contrast, although the presence of relevant values of $V_{ij}$ 
can be expected in the actual $\kappa$-type compounds 
which have large dimerization, 
the dimeric AF state in their insulating phase 
is confirmed experimentally to be stable,\cite{Kanoda} 
consistent with the MF calculations by KF without $V_{i,j}$. 
Here 
to investigate the competition between 
this dimeric AF state and the CO states, 
and to get insight into a unified view of the electronic properties 
of (ET)$_2X$, 
calculations are performed on the model in Fig. \ref{fictitious}(b) 
by varying $t_{p1}$ or $t_{p4}$.  

\begin{figure}
\begin{center}
\leavevmode\epsfysize=9cm
\epsfbox{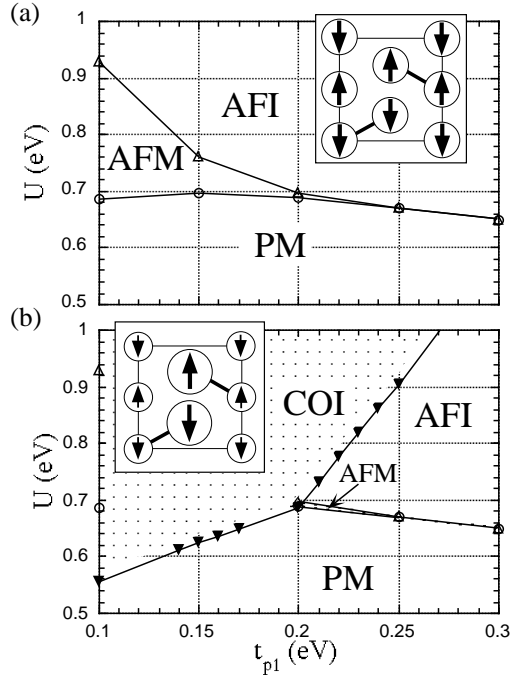}
\end{center}
\caption{Phase diagrams of the model structure in Fig. \ref{fictitious}
on the plane of $U$ and $t_{p1}$, 
in the case of $t_{p4}=t_{p}=0.1$ eV for $V=0$ (a) 
and for $V/U=0.25$ (b).
COI, AFI, AFM and PM denote the charge ordered insulating 
, antiferromagnetic insulating, antiferromagnetic metallic and paramagnetic 
metallic phases, respectively.}
\label{onlyUtb1}
\end{figure}

Only the case for the isotropic intersite Coulomb interaction $V$ 
is considered 
and the values of $t_c$ and $t_p$ are fixed at $0.01$ eV and $0.1$ eV, 
respectively.   
We just take into account a vertical stripe CO state 
described in the inset of Fig. \ref{onlyUtb1}. 
Other stripe-type solutions may have lower energy in some parameter region 
but the qualitative features on 
the competition between the stripe-type CO state and other states, 
whose investigation is our aim, 
should be made. 
Note that we know from \S \ref{theta} that 
this state is stable in the case of 
$t_{p1} = t_{p4} = t_p$, 
identical to the $\theta$-type structure with $t_p=0.1$ eV and $t_c=0.01$ eV, 
and $V/U \leq 0.49$.  
\begin{figure}
\begin{center}
\leavevmode\epsfysize=9cm
\epsfbox{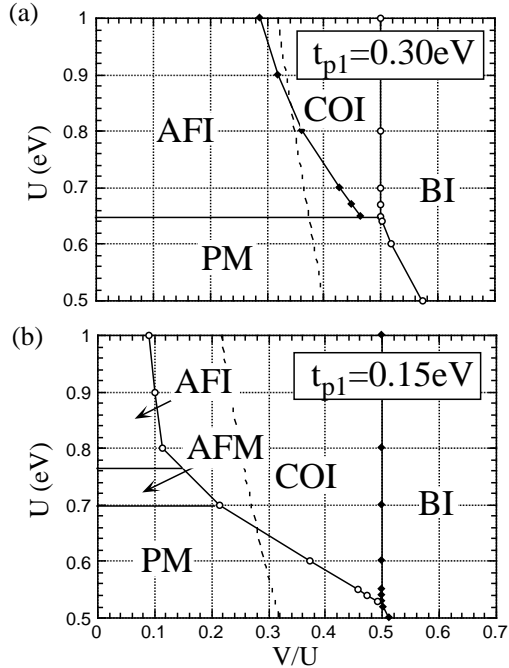}
\end{center}
\caption{Phase diagrams of the model structure in Fig. \ref{fictitious}
in the plane of $U$ and $V/U$, 
for the case of $t_{p1}=0.30$ eV (a) and $t_{p1}=0.15$ eV (b), 
where $t_c$ and $t_{p4}=t_{p}$ are fixed at 0.01 eV and 0.1 eV, 
respectively.
The notations for COI, AFI, AFM and PM are the same 
as in Fig. \ref{onlyUtb1}, 
and BI denotes the bipolaronic insulating phase. 
The dotted lines show the relation 
$2t_{p1}+U/2(1-\sqrt{1+(4t_{p1}/U)^2}) = V$, 
which is discussed in \S \ref{modeldis}}
\label{phasetb1}
\end{figure}
\begin{figure}
\begin{center}
\leavevmode\epsfysize=14cm
\epsfbox{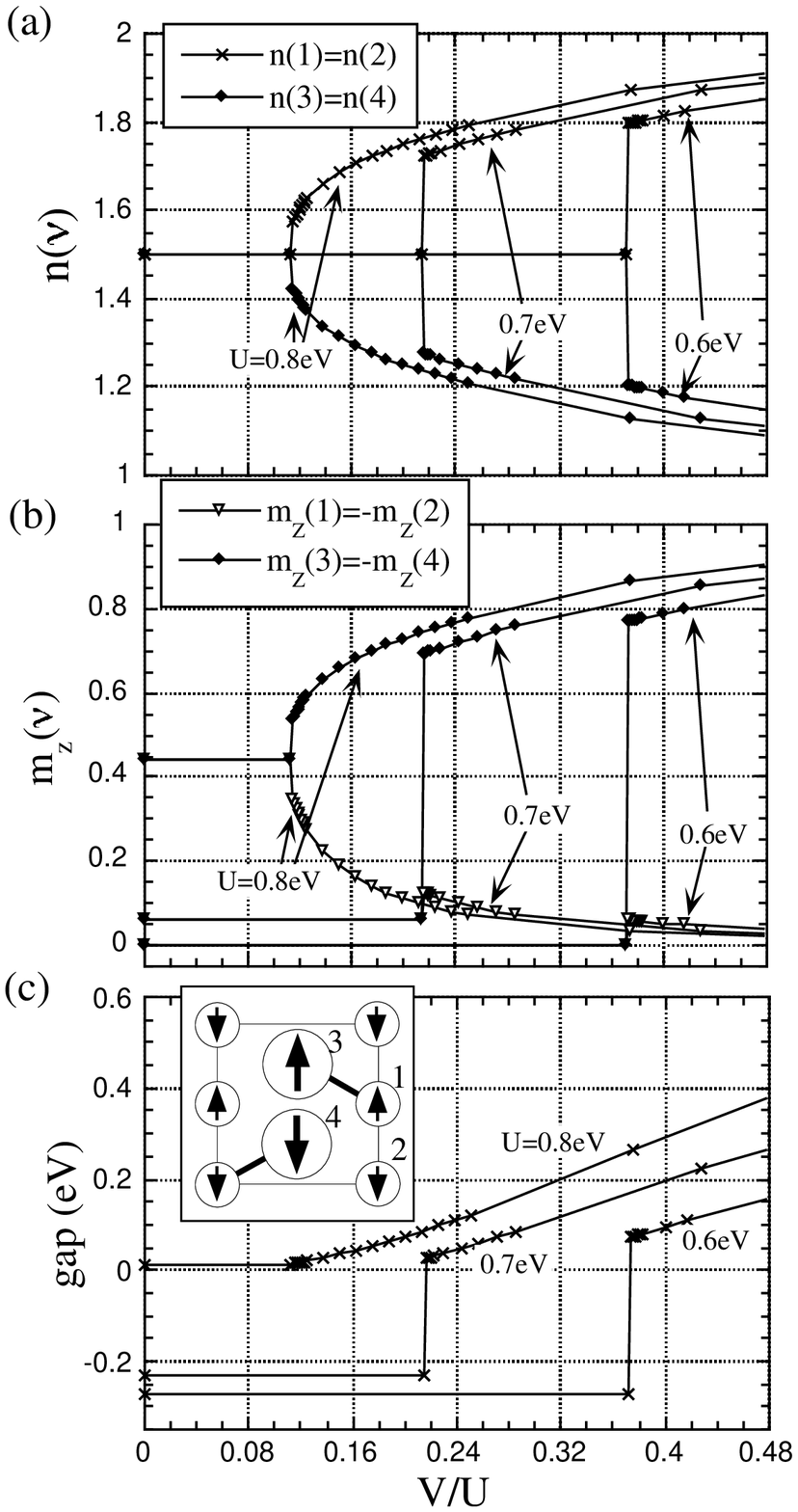}
\end{center}
\caption{$V$ dependence of the charge density (a), the magnetic moment (b)
on each site, and the band gap (c), for the case of $t_{p1}=0.15$ eV, 
$t_{p4}=t_{p}=0.1$ eV, $t_c=0.01$ eV and $U=0.8,0.7,0.6$ eV. }
\label{stripe+dimer}
\end{figure}

First 
we fix $t_{p4}=t_p$, and vary the Coulomb interactions $U$, $V$ 
and the degree of dimerization $t_{p1}$. 
The phase diagram for $V=0$ on the plane of $U$ and $t_{p1}$ 
is shown in Fig. \ref{onlyUtb1}(a), 
which is similar to the corresponding phase diagram 
for the $\kappa$-type structure obtained by KF,\cite{Kinokappa,Kinopaper} 
again suggesting that this simplified model can be considered as 
an effective model for the family of (ET)$_2X$. 
In the large $U$ region, the AF insulating state emerges 
which is the dimeric AF state, 
and becomes more stabilized as $t_{p1}$ is increased. 
\begin{figure}
\begin{center}
\leavevmode\epsfysize=8.5cm
\epsfbox{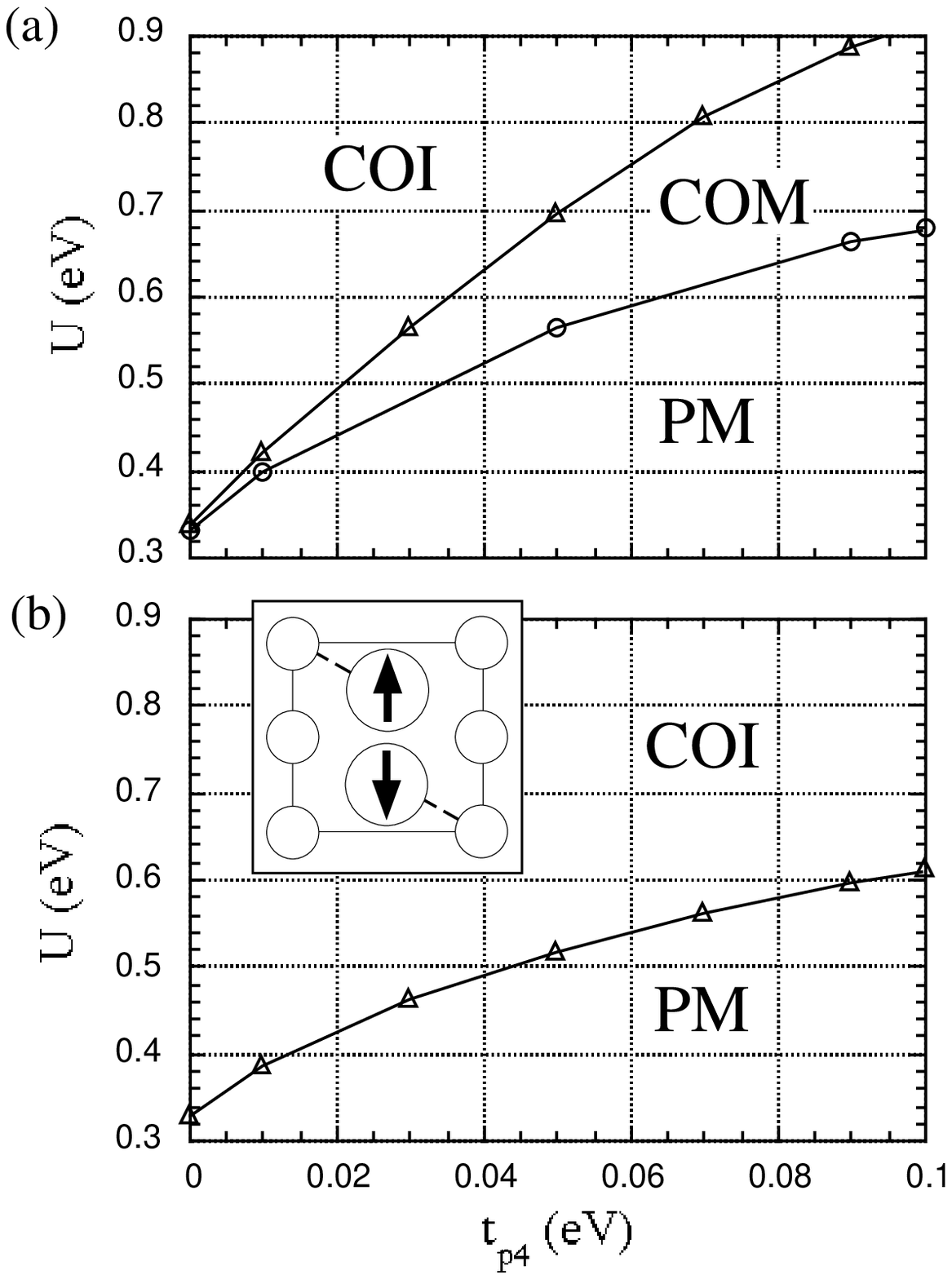}
\end{center}
\caption{Phase diagrams of the model structure in Fig. \ref{fictitious}
on the plane of $U$ and $t_{p4}$, 
in the case of $t_{p1}=t_{p}=0.1$ eV for $V=0$ (a) 
and for $V/U=0.25$ (b).
COI, COM and PM denote the charge ordered insulating, 
charge ordered metallic and paramagnetic 
metallic phases, respectively.}
\label{onlyUtb4}
\end{figure}
\begin{figure}
\begin{center}
\leavevmode\epsfysize=8.5cm
\epsfbox{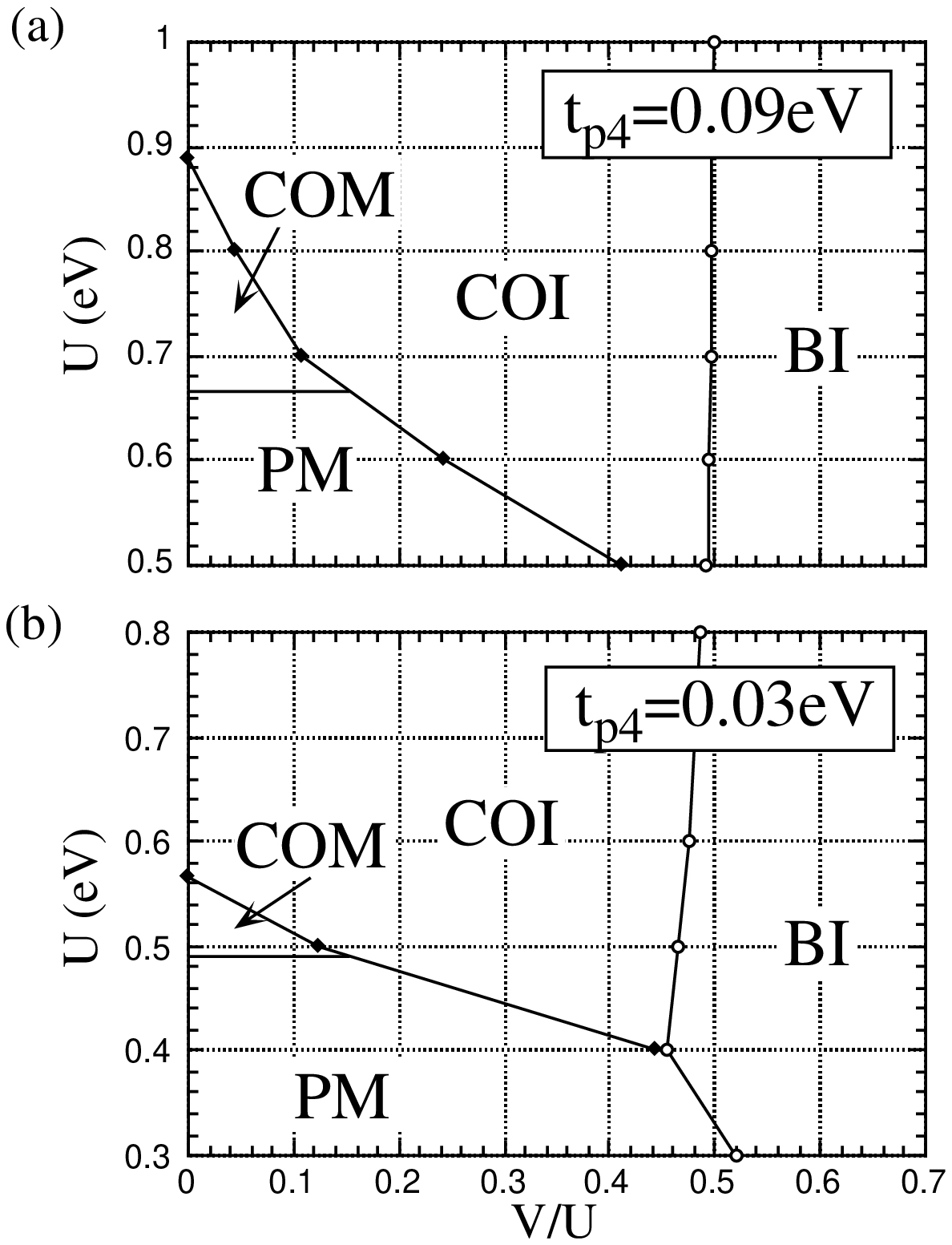}
\end{center}
\caption{Phase diagrams of the model structure in Fig. \ref{fictitious}
in the plane of $U$ and $V/U$, 
for the case of $t_{p4}=0.03$ eV (a) and $t_{p4}=0.09$ eV (b), 
where $t_c$ and $t_{p1}=t_{p}$ eV are fixed at 0.01 eV and 0.1 eV, 
respectively.
The abbreviations for COI, COM and PM are same 
as in Fig. \protect\ref{onlyUtb4}, 
and BI stands for the bipolaronic insulating phase. }
\label{phasetb4}
\end{figure}
When $V$ is turned on, there appears the CO phase. 
The phase diagrams on the plane of $U$ and $V/U$ for $t_{p1}=0.30$ eV
and $t_{p1}=0.15$ eV are shown 
in Fig. \ref{phasetb1}(a) and (b), respectively. 
It can be seen that the stripe-type CO phase is stabilized 
when $V$ exceeds some critical value, 
{\it e.g.} 0.150 eV for $t_{p1}=0.15$ eV and $U=0.7$. 
The bipolaronic state has the lowest energy 
when $V$ is larger than approximately $U/2$, 
similar to the results in \S \ref{theta}-\ref{alpha}. 
If we consider the stripe-type CO state 
with 1 hole on each site along the stripes 
and the bipolaronic state with 2 holes on every 4 sites 
which will be stabilized in the limit of $U,V \rightarrow \infty$, 
the energy loss in the Coulomb interactions 
in the stripe-type CO state and the bipolaronic CO state will be 
$2V$ and $U$ per unit cell of four sites, respectively, 
so the critical value of $V \simeq U/2$ can be easily understood. 
As can be seen in Fig. \ref{stripe+dimer}, 
where the charge density and the magnitude of spin density  
at each site together with the band gap are displayed for $t_{p1}=0.15$ eV, 
the phase transition from the AF insulating phase 
to the CO insulating phase at $V_{\rm cr1}$ is a second-order transition for 
$U=0.8$ eV, whereas the one 
from the metallic phases to the CO insulating phase 
is a first-order one for $U=0.7$ eV and 0.6 eV. 
The stripe-type CO phase which has appeared by the inclusion of $V$ can 
be described on the plane of $U$ and $t_{p1}$ as in Fig. \ref{onlyUtb1}(b), 
where the phase diagram in the case of fixed ratio of $V/U=0.25$ is shown. 
It can be seen that for $t_{p1} \lsim 0.2$ eV, 
where the dimerization is small, 
the dimeric AF phase is overwhelmed by the stripe-type CO phase, 
whereas when the dimerization is large enough 
as $t_{p1} \gsim 0.2$ eV this dimeric AF phase survives. 
\begin{figure}
\begin{center}
\leavevmode\epsfysize=15cm
\epsfbox{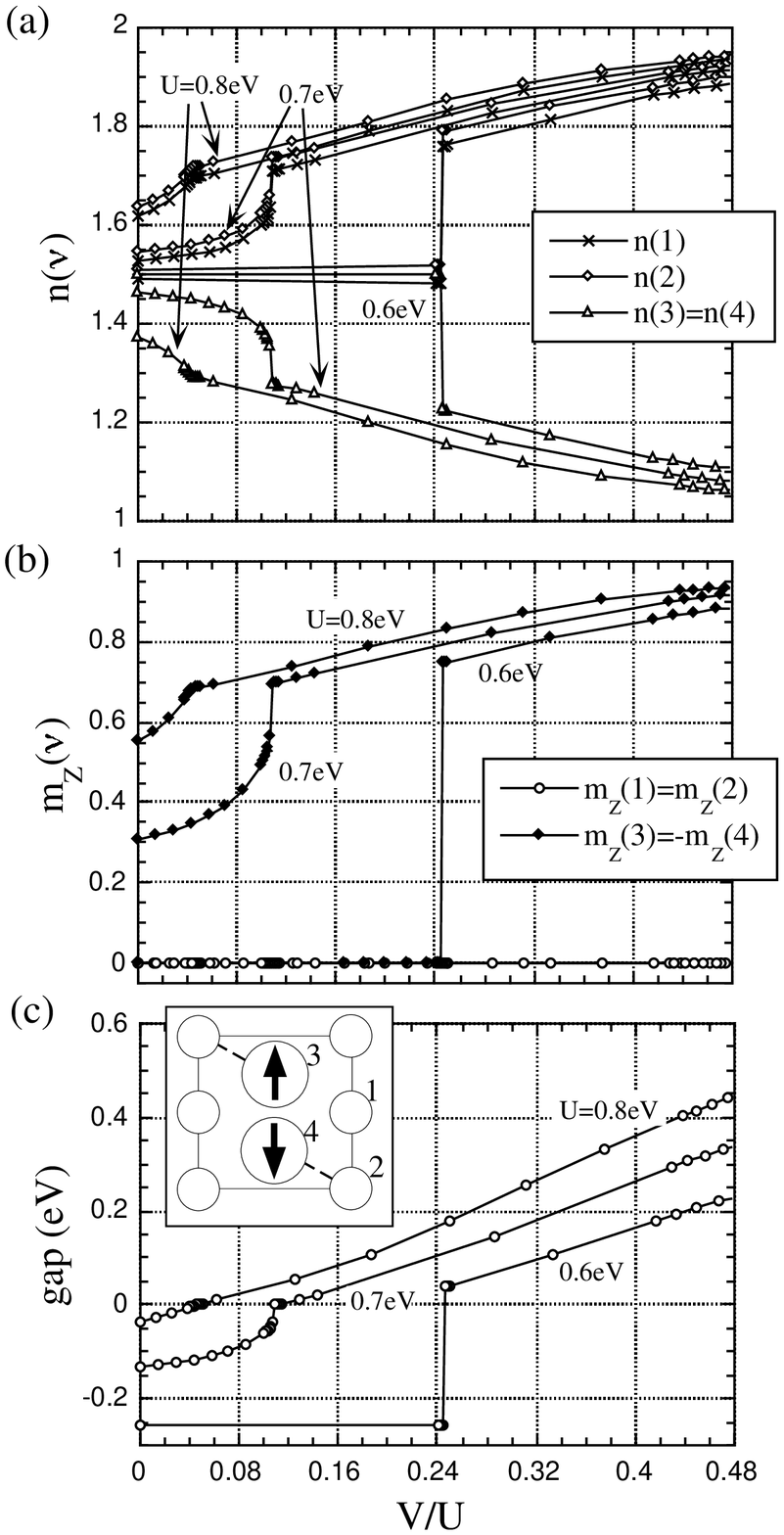}
\end{center}
\caption{$V$ dependence of the charge density (a), the magnetic moment (b)
on each site, and the band gap (c), for $t_c=0.01$ eV, 
$t_{p1}=t_{p}=0.1$ eV, $t_{p4}=0.09$ eV and $U=0.8,0.7,0.6$ eV.}
\label{stripetb4}
\end{figure}

Next we fix $t_{p1}=t_p$, and vary $t_{p4}$, 
so that the band overlap can be controlled, 
corresponding to the case of $\alpha$-type compounds. 
The phase diagram for $V=0$ on the plane of $U$ and $t_{p4}$ 
is shown in Fig. \ref{onlyUtb4}, 
where the general features are similar to the Hartree-Fock phase diagram 
for the $\alpha$-type structure obtained by KF,\cite{Kinoalpha,Kinopaper} 
as expected. 
The decrease in the value of $t_{p4}$ 
leads to the stabilization of the CO state, 
where the charge pattern and the spin configuration are schematically shown in 
the inset of Fig. \ref{onlyUtb4}(b). 
By the inclusion of $V$ the stripe-type CO phase is stabilized further. 
Phase diagrams on the plane of $U$ and $V/U$ 
for the case of $t_{p4}=0.03$ eV and 0.09 eV 
are shown in Fig. \ref{phasetb4}(a) and (b), respectively. 
It can be seen that the stripe-type CO phase is stabilized 
in a wide range of parameters, 
while the bipolaronic state again has lower energy for $V \gsim U/2$. 
In Fig. \ref{stripetb4}, 
the charge density and the magnitude of spin density  
at each site together with the band gap are shown 
for the choice of $t_{p4}=0.09$ eV. 
Here the phase transition as a function of $V$ is of second-order-type 
when it is from the CO metallic phase to the CO insulating phase 
for $U=0.8$ eV and 0.7 eV, 
and the one from the PM phase to the CO insulating phase 
is a first-order one for $U=0.6$ eV. 
The phase diagram on the plane of $U$ and $t_{p4}$ 
for the fixed ratio of $V/U=0.25$ is shown in Fig. \ref{onlyUtb4}(b), 
where the CO phase extends compared to the case of $V=0$, 
especially in the region of large $t_{p4}$.

\section{Discussions}\label{discussions}
In the previous section, it was seen that 
stripe-type CO states are stabilized in general, 
while the actual spatial pattern of charge density 
sensitively depends on the parameters of models, 
i.e. the anisotropy of the transfer integrals 
and the values of intersite Coulomb energies. 
In order to find out whether these CO states are realized 
in the actual compounds,  
and, if they are, which type of stripes are realized, 
the comparison between our calculation 
and the properties observed in experiments 
will be pursued in the following, 
since the value of the intersite Coulomb energies $V_{i,j}$ 
as well as its degree of anisotropy $V_p/V_c$ is 
difficult to determine by theories above. 
In pursuing this, however, 
we adopt from the studies\cite{Ducasse} mentioned in \S \ref{intro}
the ratio of the neighboring Coulomb energies $V_c, V_p$ 
to the on-site one $U$ 
to be $V_c/U, V_p/U \simeq 0.2-0.5$ 
and $V_p/V_c$ to be about 1.  

To reconcile the results of the MF calculations 
showing ordering of spins 
to the observed properties where 
magnetically ordered ground states are frequently destroyed 
by quantum fluctuation which is neglected in our approximation, 
the effects of this quantum fluctuation is incorporated 
by mapping the stripe-type CO states 
to the $S=1/2$ Heisenberg Hamiltonians where these spins are interacting 
by the superexchange interactions $J_{i,j}$ 
deduced from the transfer integrals in Fig. \ref{structure}.\cite{SeoBETS} 
Although 
the estimation of the values of $J_{i,j}$ along the stripes 
will be done in the following 
by using the relation $J_{i,j}\sim 4t_{i,j}^2/U$ with $U=0.7$ eV, 
we note that the definite values should not be taken seriously, 
but only the relative values should, 
since there is ambiguity on both the value of $U$ 
and the transfer integrals 
based on the simple semi-emperical extended H{\" u}ckel method. 
However, as a general trend, 
the exchange couplings get reduced 
from these values of $J_{i,j}\sim 4t_{i,j}^2/U$
in the charge ordered state 
with intermediate charge disproportionation,\cite{Thalmeier} 
thus it is natural that the exchange couplings deduced from 
experiments are smaller than those naively estimated 
from this relation. 

\subsection{$\theta$-phase}

The general features in the phase diagram in Fig. \ref{phasetheta} 
can be understood as follows. 
In the region of small $V$, 
the holes tend to form stripes along the direction 
with smaller transfer energy $t_c$, 
i.e. the vertical stripes, 
so that they can gain kinetic energy along the bonds with $t_p$. 
On the other hand,  
the holes are expected to have 
much localized character in the region of large $V$,  
then to gain exchange energy along the stripes 
they choose the stripes with larger exchange energy $J_p \sim 4t_p^2/U$, 
i.e. the diagonal stripes along $t_p$. 

When $t_c=0$, 
the topology of our model is identical to that of the 2D square lattice 
model with nearest neighbor transfer integral $t=t_p$. 
The results above for small $V$ 
that the vertical stripes along the smaller transfer integral $t=t_c$ 
are stable, 
is similar to the results in 
theoretical studies on the Hubbard model on such 2D square lattice, 
where the stability of stripes along the bonds for $U/|t| \lsim 4$ 
and those along the diagonal direction of the bonds 
for $U/|t| \gsim 4$ has been indicated,\cite{stripeson2dhm} 
since our calculations were performed for the value of $U/|t_p| = 7$, 
which correspond to the latter case, 
and the vertical stripes in our notation are 
along the diagonal direction of the bonds 
with the larger transfer integral $t_p$, 
although the filling of $n=1/4$ in our calculations 
is different from the above studies 
for $n=1/2-x$ with $x \sim 0$. 

From the MF calculations, the candidates for the ground state of 
the insulating phase in actual $\theta$-type compounds 
are the CO states with vertical stripes and those with diagonal stripes. 
The bipolaronic state is not likely realized 
since the ground state in these compounds are magnetic, 
as disclosed by the magnetic susceptibility measurement.\cite{Mori} 
When the CO states with the vertical stripes and the diagonal stripes 
are mapped to 1D $S=1/2$ Heisenberg models, 
the AF couplings between the localized spins
can be deduced to be the order of 
$J_c \sim 4t_c^2/U = 10$ K and $J_p \sim 4t_p^2/U = 500$ K, 
respectively. 
The magnetic susceptibilities measured in experiments 
for $\theta$-(ET)$_2$RbZn(SCN)$_4$ with rapid cooling condition and 
for $\theta$-(ET)$_2$CsZn(SCN)$_4$
show Curie-like behaviors,\cite{Mori,Nakamura} 
suggesting small values of exchange coupling between localized spins, 
which may indicate the vertical stripes, 
also inferred from the reflectivity measurement 
on $\theta$-(ET)$_2$RbZn(SCN)$_4$ and $\theta$-(ET)$_2$RbCo(SCN)$_4$
with rapid cooling condition.\cite{Tajima}
On the other hand 
the NMR data on $\theta$-(ET)$_2$RbZn(SCN)$_4$ with rapid cooling condition 
suggest that the charge density on molecules 
becomes continuously distributed by decreasing temperature,\cite{Miyagawa3} 
as in the case of the incommensurate CDW state. 
This is in contrast with situation in the $\theta_d$-phase 
discussed in the next subsection 
and apparently inconsistent with our stripe-type CO state 
suggesting only two kinds of molucules with different charge densities. 
However, since there are two degenerate CO states with vertical stripes  
in the $\theta$-type structure, 
the dynamical motions of stripes may exist down to low temperature, 
making difficult to detect the stripe-type CO state by NMR experiments. 
This charge fluctuation may also be the origin for the 
apparently small couplings between spins, 
inferred from the magnetic susceptibility measurements.
The interchain couplings between the spin on these vertical stripes 
show frustration due to the structure (see Fig. \ref{structure}(b)), 
which is consistent with the absence of 
AF longe range order down to 2 K.\cite{Nakamura} 

Nevertheless, a recent X-ray study on another member of this family, 
$\theta$-(ET)$_2$CsCo(SCN)$_4$,\cite{Nogami} 
shows a diffusive spot at $(0,0,1/2)$ below 20 K 
which is proposed to be due to the existence of short range CO 
along the $c$-axis with period of 2 ET molecules, 
which is the case of the horizontal stripes or the diagonal stripes. 
The discrepancy between this fact 
and the conclusion above of the existence of CO state with vertical stripes 
in $\theta$-type compounds based on experimental results 
on members with $MM'=$ RbZn, RbCo and CsZn, 
may be due to the tendency toward structural instability in 
$\theta$-(ET)$_2$CsCo(SCN)$_4$, 
i.e. toward the $\theta_d$-type structure 
as in $\theta$-(ET)$_2$RbZn(SCN)$_4$, 
and the fact that X-ray experiments are 
more sensitive to the lattice distortion 
than to the charge disproportionation. 
This possibility has also been pointed out 
by T. Mori {\it et al} from thermoelectric power measurements.\cite{TMoriCsCo}
Another possibility is that the charge pattern of the stripes is different 
between the members of $\theta$-(ET)$_2MM'$(SCN)$_4$ 
with $MM'=$ RbZn, RbCo and CsZn, 
and the salt with $MM'=$ CsCo, 
although their ET layers are isostructural.  

\subsection{$\theta_d$-phase}

The stability of the CO state with horizontal stripes 
in our MF calculations for the $\theta_d$-type structure 
compared to the results for the $\theta$-type one 
where the horizontal stripe solutions have rather high energies, 
may be related to the fact that the PM state 
in the $\theta_d$-type structure, 
which is the ground state for small $U$ and $V$, 
shows slight charge disproportionation with pattern of charge density   
similar to that of the horizontal stripe-type CO state 
described in Fig. \ref{en-Vp_thd_U07V025}. 

There are several candidates for the CO state 
realized in the insulating phase of the actual compounds in $\theta_d$-phase 
based on our calculations. 
Among them it can be deduced that the vertical stripes 
are not realized in $\theta_d$-(ET)$_2$RbZn(SCN)$_4$ since 
the temperature dependence of the magnetic suscpetibility data 
for 15 K $\lsim T \leq T_{\rm str}=195$ K 
suggest a larger exchange coupling, $J\sim $ 160K,\cite{Mori,comment}
than those inferred from the vertical stripes, 
$J_{c1} \sim 4t_{c1}^2/U = 50$ K and $J_{c2} \sim 4t_{c2}^2/U = 100$ K. 
From this experimental fact, it is clear that 
the bipolaronic state with no spins again can be excluded. 
The remaining candidates are the diagonal stripes 
and the horizontal stripes along $t_{p4}$ 
as shown in Fig. \ref{en-Vp_thd_U07V025}. 
The CO state with diagonal stripes leads to 
Heisenberg chains of four kinds of couplings,  
$J_{p1}-J_{p4}$, 
which may have a spin gap at the ground state but 
inconsistent with the Bonner-Fisher behavior 
in the magnetic susceptibility data for 15 K $\lsim T \leq T_{\rm str}$ 
mentioned above. 
The CO state with horizontal stripes along $t_{p4}$, 
which we consider to be most likely, 
is mapped to Heisenberg $S=1/2$ chains with a uniform exchange coupling 
$J_{p4} \sim 4t_{p4}^2/U = 450$ K, 
which is consistent with 
this Bonner-Fisher behavior 
and also with a reflectivity measurement suggesting the existence of the 
horizontal stripes in this compound.\cite{Tajima}
If this charge pattern is realized, the spin gap behavior 
seen in experiments\cite{Mori,Nakamura,Miyagawa}
is probably due to the spin-Peierls transition along these chains. 
Note that the the lattice distortion expected 
in the spin-Peierls transition for this case does not produce superlattices 
since the unit cell along the $b$-direction initially 
contains two ET molecules in the $\theta_d$-type structure. 
The reflectivity data on the $\theta_d$-phase of 
another member (ET)$_2$RbCo(SCN)$_4$ with slowly cooling condition 
also suggest the horizontal stripes,\cite{Tajima}
thus it may be concluded that the $\theta_d$-type structure in general 
favors the horizontal stripes. 

Finally we comment on the recent experiment 
by Miyagawa {\it et al}\cite{Miyagawa} 
who have performed a careful $^{13}$C-NMR measurement 
on $\theta$-(ET)$_2$RbZn(SCN)$_4$ on the condition of slowly cooling, 
where the $\theta_d$-type structure is observed below $T_{\rm str}$. 
Their conclusion of the existence of 
two kinds of ETs with equal population in the ground state 
is consistent with our conclusion 
that the horizontal stripes along $t_{p4}$ are realized in this compound, 
since in this horizontal stripe solution 
the charge density $n(\nu)$ on the crystallographically equivalent molecules 
`1' and `4', and `2' and `3' (see Fig. \ref{structure}) 
are respectively equal, 
i.e. $n(1)=n(4)$ and $n(2)=n(3)$, 
which is not the case in other candidates mentioned above 
showing more than four kinds of ETs 
with different charge densities ({\it e.g.} see Fig. \ref{nSz-V_thd}). 
Their data showing a broad line 
and a sharp Pake doublet, ascribed to to the molecules 
with rich and less charge density, respectively, at 
$15$ K $\leq T \lsim T_{\rm str}$ seem to suggest 
that the charge fluctuation along the stripes are large 
while in the transverse direction the stripes are `pinned', 
which may explain why the resistivity shows a sudden 
increase at $T_{\rm str}$ as lowering the temperature. 
At around the temperature where the broad line turns 
into a clear Pake doublet ($\sim$10 K)
the magnetic susceptibility also shows the drop.
If it is the spin-Peierls transition temperature, 
this result implies that the lattice distortion along the stripes 
rapidly suppress the charge fluctuation. 

Here we have not discussed the physical origin of 
the structural phase transition at $T=T_{\rm str}$, 
since the lattice degree of freedom was not included in our calculations. 
However, it is possible that the emergence of 
the CO state with horizontal stripes 
and the lattice distortion from $\theta$-type to $\theta_d$-type structure 
occur cooperatively, 
as been discussed in La based high-$T_{\rm c}$ Cuprates, 
in the context of the relation between the stripe formation 
and the low temperture tetragonal lattice distortion.\cite{Yamase} 

\subsection{$\alpha$-(ET)$_2$I$_3$ and $\alpha$-(ET)$_2M$Hg(SCN)$_4$}
The difference between results 
for the $\alpha$I$_3$ and $\alpha M$Hg-type structures
concerning the charge patterns in the stable horizontal stripe solution 
for $(V/U)_{\rm cr2} \leq V/U \leq (V/U)_{\rm cr3}$
can be understood when the exchange energies 
along the stripes, $J_{p1-p4} \sim 4t_{p1-p4}^2/U$, are considered. 
The two horizontal stripes 
are those along $t_{p2}$ and $t_{p3}$ and those along $t_{p1}$ and $t_{p4}$, 
lead to gains in exchange energies of 
$J_{p2} \sim 4t_{p2}^2/U$ and $J_{p3} \sim 4t_{p3}^2/U$, 
and $J_{p1} \sim 4t_{p1}^2/U$ and $J_{p4} \sim 4t_{p4}^2/U$, respectively. 
In the $\alpha$I$_3$-type structure, 
the value of $|t_{p4}|$ is quite small compared to $|t_{p1-3}|$ 
resulting in $J_{p1}+J_{p4}<J_{p2}+J_{p3}$, 
thus the calculations showing the stability of 
the horizontal stripes along $t_{p2}$ and $t_{p3}$ 
with larger gain of exchange energies can be explained. 
On the other hand, in the $\alpha M$Hg-type structure
the values of $J_{p1-p4}$ are similar 
but $J_{p1}+J_{p4}$ is slightly larger than $J_{p2}+J_{p3}$, 
which is consistent with the calculations indicating 
the horizontal stripes along $t_{p1}$ and $t_{p4}$ 
and the calculated energies of the two CO states 
with horizontal stripes are close to each other. 

\begin{figure}[t]
\begin{center}
\leavevmode\epsfysize=5cm
\epsfbox{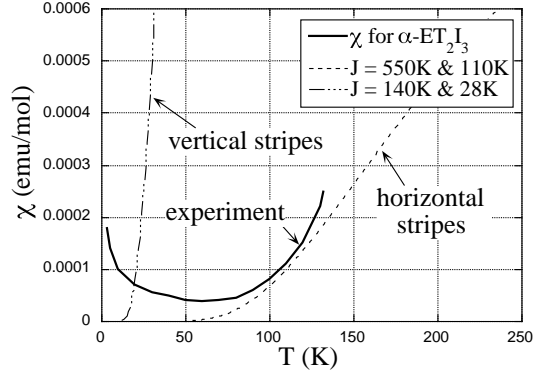}
\end{center}
\caption{Calculated magnetic susceptibility\protect\cite{Bulaevskii} 
for the Heisenberg chain models and 
the measured spin susceptibility of $\alpha$-(ET)$_2$I$_3$ taken 
from ref. \protect\ref{Rothamael}. 
}
\label{spingap}
\end{figure}

An explanation we propose here 
for the observed properties in $\alpha$-(ET)$_2$I$_3$ 
is that the stripe-type CO state is realized below $T_{\rm MI}$, 
though the charge pattern is different 
from the one proposed by KF 
which is the vertical stripes on stack I.\cite{Kinoalpha} 
The candidates for its ground state based on our 
MF calculations are two kinds of vertical stripe states, 
a horizontal stripe state and the bipolaronic state, 
as shown in Fig. \ref{en-V_al_U07}. 
The bipolaronic state may explain the experimental results 
since a nonmagnetic behavior is seen in the susceptibility measurements, 
though the required value of $V/U \gsim (V/U)_{\rm cr3}=0.54$ 
in our calculation is rather large. 
The estimations of exchange coupling along the stripes 
for the stripe-type CO states lead to uniform $J_{c1} \sim 4t_{c1}^2/U =50$ K
in the case of the vertical stripes along stack II, 
and to alternating 
$J_{c2} \sim 4t_{c2}^2/U = 140$ K and $J_{c3} \sim 4t_{c2}^2/U =30$ K 
in the case of those along stack I. 
The CO state we propose here to be realized in $\alpha$-(ET)$_2$I$_3$ 
is the horizontal stripes along $t_{p2}$ and $t_{p3}$, 
with alternating exchange couplings 
$J_{p2} \sim 4t_{p2}^2/U =1100$ K and $J_{p3} \sim 4t_{p3}^2/U = 220$ K,  
since the observed magnetic susceptibility \cite{Rothamael} below $T_{\rm MI}$ 
agree well with the calculated susceptibility for the 
alternating Heisenberg chain with 550 K and 110 K, 
which is respectively half of the the values $J_{p2}$ and $J_{p3}$, 
as compared to the one for the 
Heisenberg chains mapped from the CO state with vertical stripes, 
as shown in Fig. \ref{spingap}. 
%
%

The effect of applying pressure observed in this compound 
can be inferred from the results of calculations 
on the simplified model in \S \ref{model}, as follows. 
The pressure makes the distances of molecules shorter so that 
the values of transfer integrals increase 
while it hardly affects the on-site Coulomb energy $U$, 
so the effective on-site Coulomb interaction i.e. $U/W$ get reduced, 
where $W$ denotes the band width.
In Figs. \ref{onlyUtb4} and \ref{phasetb4},  
it can be seen that the PM phase is 
next to the CO phase in the regionof smaller $U/W$, 
which is stabilized under high pressure in $\alpha$-(ET)$_2$I$_3$, 
as has been discussed by KF.\cite{Kinopaper} 
Although the transport measurements by applying pressure and/or
magnetic field have been intensively done in this compound,\cite{Kajita} 
experiments on the magnetic properties such as NMR are crucially needed 
to understand their electronic properties, 
since in the CO state the charge and the spin degree of freedoms 
are strongly related to each other.
Note that 
it is difficult to decide whether $V_{i,j}/W$ increases or decreases 
by applying pressure to the ET salts 
since it is natural that the values of $V_{i,j}$ increase 
by applying pressure as in the case of the tranfer integrals $t_{i,j}$ 
and the anisotropy of the ET molecules 
leads to the difficulty in deducing the 
the values of intersite Coulomb energies 
$V_{i,j}$ between ET molecules, 
although the values of transfer integrals between ETs 
are carefully calculated by the extended H{\" u}ckel method 
for the cases of different configurations.\cite{TMoriGenealogy}

As for $\alpha$-(ET)$_2M$Hg(SCN)$_4$, 
the CO metallic phase found in the calculation for the 
simplified model in \S \ref{model} may be the candidate 
for the ground state 
supporting the existence of a charge disproportionated state 
proposed by some groups.\cite{MHgCDW}
However, in \S \ref{alpha} the results on the calculations 
for the actual $\alpha M$Hg-type structure, 
only insulating states with CO are found to be stable 
though the band gap for $V \sim 0.2$ has a quite small value of 
$\sim 0.05$ eV in the CO state with vertical stripes 
as seen in Fig. \ref{nSz-V_MHg_U07}. 
So at this point incommensurate SDW state 
due to the nesting of the Fermi surface, 
which is not treated in our formulation only considering 
the states commensurate to the lattice, 
is the candidate for the ground state as discussed in KF.\cite{Kinointer} 
To understand their electronic properties, 
more studies on this series is needed. 

\subsection{Unified view of (ET)$_2X$}\label{modeldis}

\begin{figure}
\begin{center}
\leavevmode\epsfysize=10.5cm
\epsfbox{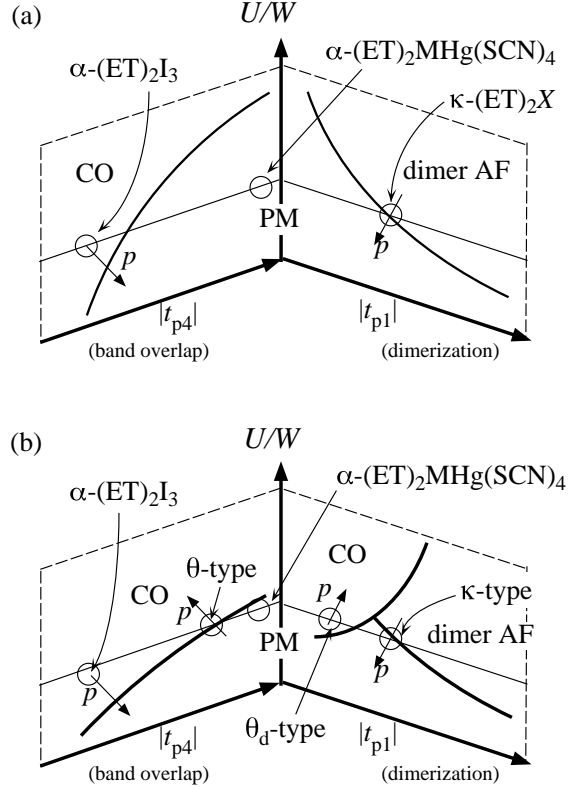}
\end{center}
\caption{Schematic phase diagrams for the unifed view of (ET)$_2X$ 
in cases of $V=0$ (a) by KF\cite{Kinopaper} 
and $V \neq 0$ (b) by the present work. 
The notations AF$\kappa$ and AF$\alpha$ are adopted from ref. \ref{Kinopaper}, 
which represents the AF phases found in their Hartree-Fock calculation
on $\kappa$ and $\alpha$-type structures, respectively. 
PM and CO denotes the paramagnetic metallic and charge ordered phases, 
respectively. 
The effect of pressure are shown by arrows. 
}
\label{phase3D}
\end{figure}

In the work of KF,\cite{Kinopaper}a unified view of (ET)$_2X$ was given 
as shown in Fig. \ref{phase3D}(a), 
where the two planes are 
based on Hartree-Fock calculations considering 
$\kappa$ and $\alpha$-type structures on appropriate Hubbard models. 
Based on the calculations in \S \ref{model} 
showing that the inclusion of intersite Coulomb interactions 
gives rise to a wide region of CO state, 
their phase diagram can be modified 
to the one as shown in Fig. \ref{phase3D}(b), 
where the two planes are 
based on the results in Figs. \ref{onlyUtb1}(b) and \ref{onlyUtb4}(b). 
The location of each type of compound is also indicated 
in Fig. \ref{phase3D}(b): 
$\kappa$-phase is in the region of large dimerization 
where the dimer AF state is stabilized, 
in contrast with $\theta_d$-phase which is located in 
in the region of small dimerization where the CO state is stable. 
$\theta$-phase is located between $\alpha$I$_3$ and $\alpha M$Hg-phases 
on the axis of band overlap. 
The main difference in the plane of band overlap 
between Figs. \ref{onlyUtb1}(b) and \ref{onlyUtb4}(b)
is that the $\alpha M$Hg-phase is located 
in the PM phase far from the CO phase in  Fig. \ref{phase3D}(a)
whereas it is close to the boundary 
between the CO phase and the PM phase in Fig. \ref{phase3D}(b), 
so the criticality toward the CO phase may play some crucial role, 
which is beyond the present study. 

The effects of pressure are indicated by arrows in Fig. \ref{phase3D}(b), 
following the discussion of Kanoda\cite{Kanoda} and KF\cite{Kinopaper} 
for the $\kappa$ and $\alpha$-phases,
and that of H. Mori {\it et al}\cite{Mori} for 
the $\theta$ and $\theta_d$-phases, respectively. 
The directions of arrows for the $\alpha$ and $\kappa$-phases 
are opposite from those for the $\theta$ and $\theta_d$-phases,  
since in the former compounds it is known that 
the application of external pressure increases the bandwidth, 
namely reduces the correlation effect,\cite{Kinopaper,Kanoda} 
while in the latter case it widens $\phi$ 
and results in the decrease of the bandwidth.\cite{Mori} 
From this point of view, it can be expected that applying pressure to the 
$\theta$-type compounds which is metallic down to low temperature 
such as $\theta$-(ET)$_2$I$_3$ 
will give birth to an insulating ground state with stripe-type CO state. 
The effect of uniaxial pressure\cite{Kagoshima} is also an intriguing method 
since it can tune the anisotropy of the system 
and the stripe-type CO states found in the present study 
must be highly sensitive to the anisotropy. 

There are discussions that 
in the case of strongly dimerized structures as in the $\kappa$-phase 
the effective on-dimer Coulomb energy can be estimated as 
$U_{\rm dimer} \simeq 2|t_{p1}|+U/2(1-\sqrt{1+(4t_{p1}/U)^2})$, 
where $t_{p1}$ is the intradimer transfer integral.\cite{Kinopaper,Kanoda,McKenzie}
$U_{\rm dimer}$ corresponds to the effective on-site Colomb energy 
when one maps models of 1/4-filling with strong dimerization 
to models with 1/2-filled band, 
{\it e.g} as has been studied recently for $\kappa$-type compounds\cite{KinoFLEX}
for the appropriate 1/2-filled Hubbard model. 
It can be said that $U_{\rm dimer}$ represent the energy scale 
in the Mott insulating state. 
On the other hand, 
as can be seen in Figs. \ref{stripe+dimer} and \ref{stripetb4}, 
it appears that the band gap 
in the CO state increases linearly when $V$ is increased, 
which is similarly seen in a classical CO system Fe$_3$O$_4$.\cite{Park,Anisimov} 
In Fig. \ref{phasetb1}(b), the line of $U_{\rm dimer} = V$ 
is drawn by the dotted line, 
which shows qualitative feature similar to the boundary between 
the CO state and the dimeric AF state 
in the case of large dimerization $t_{p1}=0.30$ eV 
and in the region of large $U$. 

\section{Summary and Conclusion}
\label{summary}
CO phenomena in 2D organic conductors (ET)$_2X$ have been investigated 
within the Hartree MF approximation on relevant extended Hubbard models 
including both the on-site and the intersite Coulomb interactions 
among which the latter are crucial to understand the CO phenomena 
in 1/4-filled systems. 
By taking into account the explicit anisotropy of
the transfer integrals and the intersite Coulomb interactions 
for the $\theta$, $\theta_d$ and $\alpha$-type structures, 
it is found that the intersite Coulomb interactions 
give rise to stripe-type CO states 
whose charge patterns are sensitively depending on the 
parameters of these models. 

Based on the results of calculations, 
implications from the existing experimetal facts
lead us that stripes along the bonds with the transfer integral $t_c$ 
are realized in $\theta$-type compounds, 
those along $t_{p4}$ in $\theta_d$-type compounds 
and those along $t_{p2}$ and $t_{p3}$ in $\alpha$-(ET)$_2$I$_3$. 
Their magnetic properties 
are well reproduced with these stripe-type CO states, 
considering the effects of quantum fluctuation 
by mapping the CO states to the $S=1/2$ Heisenberg models.
A unifed view on the electronic states in (ET)$_2X$
is obtained by taking the degree of dimerization and 
that of the band overlap as key ingredients, 
where several states compete to each other, 
which are the Mott insulating state, the stripe type CO state 
and the paramagnetic metallic state, 
realized in the actual compounds. 

Such CO states due to intersite Coulomb interaction 
are expected to be widely realized in other 
2D non 1/2-filled (non dimerized) organic compounds 
showing insulating behavior, 
{\it e.g.} $\alpha'$, $\alpha''$ and $\beta''$-phases of (ET)$_2X$
and also in (ET)$_mX_n$ with $(m,n)\neq(2,1)$ which do not give 
3/4-filled $\pi$-band, 
as well as in non ET compounds.\cite{IshiYama,TMoriGenealogy} 
The present study, which makes a guideline to understand 
the interrelationship among the experimental facts 
by taking the CO phenomena as a key factor, 
might help the understandings of their physical properties. 

\acknowledgements 
The author is grateful to Hidetoshi Fukuyama for everyday interactions. 
He also thanks H. Kohno and H. Yamase 
for enumerable fruitful suggestions, 
S. Fujiyama, K. Kanoda, K. Miyagawa, H. Mori, T. Nakamura, 
Y. Nogami, H. Tajima and T. Takahashi 
for informative discussions from the experimental point of view, 
especially K. Kanoda and K. Miyagawa for providing their preprint 
prior to publication.  
This work was supported by JSPS Research Fellowships for Young Scientists.


\begin{thebibliography}{99}
\bibitem{note}In ref. \ref{Mori}, the greek character 
$\theta$ is used to represent the dihedral angle 
but here we use $\phi$ to avoid confusion.
\bibitem{noteMHg}Signs of the transfer energies 
are modified in Fig. \ref{band1} 
from those of $\alpha$-(ET)$_2$NH$_4$Hg(SCN)$_4$, as 
$t_{p3} \rightarrow -t_{p3}$ and $t_{p4} \rightarrow -t_{p4}$, 
in order to compare the features between $\alpha$I$_3$-type and 
$\alpha M$Hg-type. 
The overall features of the band structure and the Fermi surface are 
not changed by this modification.
\bibitem{comment}In ref. \ref{Mori}, The magnetic susceptibility data 
was fitted by the 1D as well as 2D Heisenberg model, with exchange coupling 
$J=$ 157 K and 100 K, respectively. Here only the 1D case is considered 
since our MF calculation suggests the 1D ordering of charges.
\end{thebibliography}
\end{document}